\RequirePackage{fix-cm}
\documentclass[6pt]{article}
\usepackage{amsmath}\usepackage{amsmath,amsfonts,amssymb,amsthm,url,array}
\usepackage{float}
\usepackage{mathbbol}
\usepackage{abstract}
\usepackage{hyperref}
\usepackage[utf8]{inputenc}
\usepackage[T1]{fontenc}

\RequirePackage[numbers,round]{natbib}
\usepackage{setspace}
\usepackage{mathptmx}
\setcitestyle{square}
\usepackage{titlesec}
\titleformat{\paragraph}
{\normalfont\normalsize\bfseries}{\theparagraph}{1em}{}
\titlespacing*{\paragraph}
{0pt}{3.25ex plus 1ex minus .2ex}{1.5ex plus .2ex}
\date{}
\usepackage{graphicx}
\usepackage{bbm}
\usepackage{mathtools}
\usepackage{graphicx}
\usepackage{caption}
\usepackage{subcaption}
\usepackage[nomarkers,figuresonly]{endfloat}
\allowdisplaybreaks
\begin{document}
\title{Portfolio Optimization Managing Value at Risk under Heavy Tail \\ Return, using Stochastic Maximum Principle}

\author{
    Corresponding Author: Subhojit Biswas\\
    \textit{M.Tech, Indian Statistical Institute, Kolkata}\\
    \textit{subhojit1016kgp@gmail.com}\\
    \\
    \\
    Mrinal K.Ghosh\\
    \textit{Professor, Department of Mathematics}\\
    \textit{Indian Institute of Science, Bangalore}\\
    \textit{mkg@math.iisc.ernet.in}\\
    \\
    \\
    Diganta Mukherjee\\
    \textit{Professor, Sampling and Official Statistics Unit}\\
    \textit{Indian Statistical Institute, Kolkata}\\
    \textit{digantam@hotmail.com}\\
}


\maketitle
\newpage
\begin{abstract}
We consider an investor whose portfolio consists of a single risky asset and a risk free asset. The risky asset's return has a heavy tailed distribution and thus does not have higher order moments. Hence, she aims to maximize the expected utility of the portfolio defined in terms of the median return. This is done subject to managing the Value at Risk (VaR) defined in terms of a high order quantile. Recalling that the median and other quantiles always exist and appealing to the asymptotic normality of their joint distribution, we use the stochastic maximum principle to formulate the dynamic optimization problem in its full generality. The issue of non-smoothness of the objective function is  resolved by appropriate approximation technique.  We also provide detailed empirical illustration using real life data. The equations which we obtain does not have any explicit analytical solution, so for numerical work we look for accurate approximations to estimate the value function and optimal strategy. As our calibration strategy is non-parametric in nature, no prior knowledge on the form of the distribution function is needed. Our results show close concordance with financial intuition. We expect that our results will add to the arsenal of the high frequency traders.
\end{abstract}

\newpage
\section{Introduction}

\subsection{Background and Motivation}

Risk management occurs everywhere in the financial world. There are lot of places where risk management is done, such as it occurs when an investor buys low-risk government bonds over riskier corporate bonds, bank performing a credit check on an individual before issuing a personal line of credit, stockbrokers buying assets like options $\&$ futures in their portfolio and money managers using strategies like portfolio and investment diversification to mitigate or effectively manage risk. Inadequate risk management can result in severe consequences such as the sub prime mortgage meltdown of 2007 that triggered the Great Recession that stemmed from poor risk-management decisions. In the financial world the performance benchmark of the portfolio associated with the risk and portfolio management is primarily risk management. A common definition of investment risk is a deviation from an expected outcome, which we can benchmark with the market parameters. The deviation can be positive or negative. How Do Investors Measure Risk?
Investors use a variety of tactics to ascertain risk. One of the most commonly used risk metrics is Value at Risk (VaR), a statistical measure of the riskiness of financial entities or portfolios of assets. It is defined as the maximum dollar amount expected to be lost over a given time horizon, at a pre-defined confidence level. There are also other risk measure metrics used in the market such as Sharpe Ratio or Expected Shortfall (ES). Our main focus in this paper will be Value at Risk (VaR).

The particular situation that we consider here is when return of the assets follow a heavy tail distribution whose higher order moments do not exist. In such a  situation, the expected return and conventional volatility measure values may not exist. But, we know that the quantiles of the distribution always exist and we take advantage of this fact.
Our aim in this paper is to find a strategy for the investor such that the median return (as a proxy for expected return) is maximised subject to the VaR (as a measure of risk) being kept above a critical level with a high probability. Here the VaR is considered at a certain quantile level (the common choice of 5\%). The median is the 50\% quantile. There are other important constraints which are needed to be addressed in a portfolio optimisation situation. A salient one is the transaction cost. Here we also address this issue in the portfolio optimization problem and we discuss how the investors' are going to follow a recursive optimal policy such that, simultaneously, the VaR is managed at the desired level along with optimizing the median return. Thus, in this paper we have proposed a continuous time dynamic framework for the investor on how to handle the heavy tail distribution function of return. Our proposal does not require knowledge about the exact form of the distribution. We take recourse to non-parametric calibration techniques to handle general unknown distribution functions.

\subsection{Literature Review}
Interest rate risk immunization is one of the key concerns for fixed income portfolio management. In recent years, risk measures (e.g. value‐at‐risk and conditional value‐at‐risk) as tools for the formation of an optimum investment portfolio has gained traction. The article by \citeauthor{Ref5}\cite{Ref5} aims to discuss this issue. The work by \citeauthor{Ref3}\cite{Ref3} which gives an idea of portfolio optimization under drawdown constraint and stochastic Sharpe ratio tells us how the stochastic differential equation of the asset return can be converted in to the stochastic differential equation of the quantile. \citeauthor{Ref6}\cite{Ref6} aims to test empirically the performance of different models in measuring VaR and ES in the presence of heavy tails in returns using historical data. Daily returns are modelled with empirical (or historical), Gaussian and Generalized Pareto (peak over threshold (POT) technique of extreme value theory (EVT)) distributions. Assessing financial risk and portfolio optimization using a multivariate market model with returns assumed to follow a multivariate normal tempered stable distribution (i.e. this distribution is a mixture of the multivariate normal distribution and the tempered stable subordinator) can be seen in the paper of  \citeauthor{Ref7}\cite{Ref7}. Several authors have considered the optimal portfolio problem under drawdown constraint. The first to comprehensively study this problem over infinite time horizon in a market setting with single risky asset modelled as a geometric Brownian motion with constant volatility (log normal model) was \citeauthor{Ref8}\cite{Ref8}. Dynamic programming was used to solve the maximization problem of the long term growth rate of the expected utility of the wealth. \citeauthor{Ref9}\cite{Ref9} streamlined the analysis of \citeauthor{Ref8}\cite{Ref8} and extended the results to the case when there are multiple risky assets. The paper by \citeauthor{Ref10}\cite{Ref10} gives an idea of portfolio selection by stochastic dynamic programming. Finally, very relevant for our empirical analysis, we mention the paper by  \citeauthor{Ref12}\cite{Ref12} who gives an idea of using a non parametric estimator for the State Price Densities implicit in option prices. The paper by \citeauthor{Ref20}\cite{Ref20} uses the approximate dynamic programming to set up a Markov decision model for the multi-time segment portfolio with transaction cost. The paper by \citeauthor{Ref21}\cite{Ref21} analyses the contract between an entrepreneur and an investor, using a non-zero sum game in which the entrepreneur is interested in company survival and the investor in maximizing expected net present value. This paper looks into a different setup for a finance company. The paper by \citeauthor{Ref22}\cite{Ref22} analyses the data of the CSI 300 Index in the past five years, and uses Monte Carlo simulation and historical simulation to calculate the VaR of the five-year index and test its validity. It combines the result with China’s market economy and puts forward some suggestions on financial risk management in China’s financial market. The main contribution of the paper by \citeauthor{Ref23}\cite{Ref23} is to analyze the application of multi-state Markov models to evaluate credit card risk by investigating the characteristics of different state transitions in client-institution relationships over time, thereby generating score models for various purposes. It gives a different direction of application of the Markov Decision Process in the financial market. The paper by \citeauthor{Ref24}\cite{Ref24} considers the variance optimization problem of average reward in continuous-time Markov decision process (MDP). It assumes that the state space is countable and the action space is a Borel measurable space. The main purpose of the paper is to find the policy with the minimal variance in the deterministic stationary policy space. The paper by \citeauthor{Ref25}\cite{Ref25} studies a portfolio optimization problem combining a continuous-time jump market and a defaultable security and provides a numerical solutions through the conversion into a Markov decision process. The paper also analyse allocation strategies under several families of utility functions and compares it with previously obtained results. The paper by \citeauthor{Ref26}\cite{Ref26} studies a portfolio consisting of a single risky asset and a risk free asset in discrete time. The aim is to maximize the expected utility of the portfolio subject to the Value at Risk assuming a heavy tail distribution of the stock price return. It uses Markov Decision Process and dynamic programming principle to get the optimal strategies and the value function which maximize the expected utility for parametric as well as non parametric distributions. We aim to generalise this to the continuous time case which is more suitable to address the concerns of high frequency trading.

\subsection{Our Contribution}

In this article the investor is worried about when to build up on stocks or liquidate the stock when dealing with heavy tail distribution of the return of the stock prices and tries to optimize the portfolio based on Value at Risk (Var) in continuous time. The investor's portfolio has one risky asset and a risk free asset. We consider appropriate quantiles of the heavy tailed distribution as proxy for mean return and risk. We first show that these quantiles asymptotically jointly follow the multivariate normal distribution; allowing us to invoke the conventional modelling technique based on the Ito equtions. This also ensures, as in the usual continuous time set up, the self-financing conditions being successfully invoked for this formulation. Thus, we formulate and work with the stochastic differential equation for the quantiles. Then we use the stochastic maximum principle to formulate the dynamic optimisation problem following \citeauthor{Ref13}\cite{Ref13}. The equations which we obtain does not have any explicit analytical solution, so we look for suitable approximations to estimate the value function and optimal strategy. As our calibration strategy is non-parametric in nature, no prior knowledge on the form of the distribution function is needed.

\subsection{Organization of the paper}
In Section 2 we consider the quantiles of the heavy tailed distribution which are asymptotically multivariate normal with a very simple co-variance structure and a mean. The stochastic differential equation is derived for the quantiles. Subsequently we derive the stochastic maximum principle for the optimal portfolio problem under certain assumptions and present the analytical formula for the optimal portfolio strategy. Examples and numerical results are included in the subsections of Section 3. Finally Section 4 concludes.

\section{Formulation and Analysis}

We assume the existence of a friction-less financial market. In our portfolio we consider a risky asset denoted by S and a risk-free asset, such as bank account, providing a risk-free rate of interest given by a scalar constant r > 0.
Let the return of the risky asset at a time instant be given by $dX_t = \frac{dS}{S}$ where $X_t$ follows a heavy tail distribution with population c.d.f. F(x), which is assumed continuous and differentiable to at least second order. Heavy tailness of the distribution does not allow us to formulate the linear stochastic differential equation for the return of the stock. Instead, we will focus on two quantiles (as discussed above, we need to consider the median and lower 5\% quantile together for our portfolio optimisation)  $X_{(p_1)}$ and $X_{(p_2)}$, for $0 < p_1 < p_2 < 1$ (below, we will specifically consider $p_1 = 0.05$ and $p_2 = 0.5$), and take recourse to some usual asymptotics using the following proposition from \citeauthor{Ref1}\cite{Ref1}.

Let a random sample of size N be given from this population and let the observations be ordered by size from the smallest $(X_{(1)})$ to the largest $(X_{(N)})$ so that $X_{(1)}$ $\leq$ $X_{(2)}$ $\leq$.....$\leq$ $X_{(N)}$. Then let the sample quantile $\xi_{(p)}$ be defined as the r-th order statistic,  $X_{(r)}$, where r = [Np] denotes the greatest integer less than or equal to Np. If F is strictly monotonic, $\xi_{(p)}$ has the property of strong or almost sure consistency \citeauthor{Ref2}\cite{Ref2}. Note that this result does not require the existence of moments for $F(x)$, which is often the case for heavy tailed distributions. \\

\textbf{Proposition: } If F is differentiable at $\xi_p$ for p $\in$ $\left\{ p_{1}, p_2\right\}$ with density $f(\xi_{(p_1)}) = f_1$ and $f(\xi_{(p_2)}) = f_2$, then $\xi_{(p)}$'s are asymptotically multivariate normal with a simple co-variance structure. From Lemma 1 of \citeauthor{Ref1}\cite{Ref1} we can write this for two quantiles as
\[
\Lambda=
\begin{bmatrix}
    \frac{p_1(1-p_1)}{f_1^2} & \frac{p_1(1-p_2)}{f_1f_2} \\
    \frac{p_1(1-p_2)}{f_1f_2} & \frac{p_2(1-p_2)}{f_2^2}
\end{bmatrix}.
\]

On the basis of asymptotic normality, the equation of motion for the quantiles can be written as

\[
\begin{bmatrix}
    d X_{(p_1)}(t) \\
    d X_{(p_2)}(t)
\end{bmatrix} = \begin{bmatrix}
 \mu_1 \\
   \mu_2
\end{bmatrix} dt +
\begin{bmatrix}
    \frac{p_1(1-p_1)}{f_1^2} & \frac{p_1(1-p_2)}{f_1f_2} \\
    \frac{p_1(1-p_2)}{f_1f_2} & \frac{p_2(1-p_2)}{f_2^2}
\end{bmatrix}\begin{bmatrix}
    dW_t^{(1)} \\
    dW_t^{(2)}
\end{bmatrix}.
\]

\noindent where $dW_t^{(1)}$ and $dW_t^{(2)}$ are two independent Brownian motions

For reducing the number of parameters, we subtract the expected value of one of the quantiles from the data. Thus, the expected value of one of the quantile becomes zero and the values of others are actually relative to this one. Applying this methodology the equation of motion now becomes

\[
\begin{bmatrix}
    d X_{(p_1)}(t) \\
    d X_{(p_2)}(t)
\end{bmatrix} = \begin{bmatrix}
 0 \\
   b_2
\end{bmatrix} dt +
\begin{bmatrix}
    \frac{p_1(1-p_1)}{f_1^2} & \frac{p_1(1-p_2)}{f_1f_2} \\
    \frac{p_1(1-p_2)}{f_1f_2} & \frac{p_2(1-p_2)}{f_2^2}
\end{bmatrix}\begin{bmatrix}
    dW_t^{(1)} \\
    dW_t^{(2)}
\end{bmatrix}.
\]
Simplifying notation, we denote $X_{(p_1)}$ by $X_1$ and $X_{(p_2)}$ by $X_2$ to write

\[
\begin{bmatrix}
    d X_1(t) \\
    d X_2(t)
\end{bmatrix} = \begin{bmatrix}
0 \\
   b_2
\end{bmatrix} dt +
\begin{bmatrix}
    \frac{p_1(1-p_1)}{f_1^2} & \frac{p_1(1-p_2)}{f_1f_2} \\
    \frac{p_1(1-p_2)}{f_1f_2} & \frac{p_2(1-p_2)}{f_2^2}
\end{bmatrix}\begin{bmatrix}
     dW_t^{(1)} \\
    dW_t^{(2)}
\end{bmatrix}.
\]
So finally expanding,
\begin{equation}
    d X_1(t) = \frac{p_1(1-p_1)}{f_1^2}dW_t^{(1)} + \frac{p_1(1-p_2)}{f_1f_2}dW_t^{(2)}  \nonumber
\end{equation}

\begin{equation}
    d X_2(t) = b_2 dt + \frac{p_1(1-p_2)}{f_1f_2}dW_t^{(1)} + \frac{p_2(1-p_2)}{f_2^2}dW_t^{(2)}. \nonumber
\end{equation}

We denote the wealth process of an investor by $L$ who invests $\pi_t$ portion of it in risky asset and the remaining in the bank which is a risk-free asset.
 The strategy $\pi_t$ is assumed to be ${\mathcal{F}}_t$-adapted, where ${\mathcal{F}}_t$ is the filtration generated by $(W^{(1)}_t, W^{(2)}_t)$, satisfying
 $$ \int_0^T \mathbbm{E} (\pi_t)^2 \: dt \: < \infty.$$
 In this paper we evaluate the performance of the portfolio by the median of this wealth process which we denote by $L^{(1)}$. The equation of motion for $L^{(1)}$ is given by

\begin{equation}
    d L_{t}^{(1)} = L_{t}^{(1)}(1 - \pi_t) r dt + L_{t}^{(1)} \pi_t d X_1 \nonumber
\end{equation}

\begin{equation}
\textit{or, }    d L_{t}^{(1)} = L_{t}^{(1)}(1 - \pi_t) r dt + L_{t}^{(1)} \pi_t \Big(\frac{p_1(1-p_1)}{f_1^2}dW_t^{(1)} + \frac{p_1(1-p_2)}{f_1f_2}dW_t^{(2)} \Big) \nonumber
\end{equation}

\begin{equation}\label{eq1}
\textit{or, }     d L_{t}^{(1)} = L_{t}^{(1)}(1 - \pi_t) r dt + L_{t}^{(1)} \pi_t \frac{p_1(1-p_1)}{f_1^2} dW_t^{(1)} + L_{t}^{(1)} \pi_t \frac{p_1(1-p_2)}{f_1f_2} dW_t^{(2)}.
\end{equation}

In this work, we propose an investment framework that encourages managing the Value at Risk,
while maximizing the median value of the utility function U satisfying:

\textbf{Assumption 1. }The terminal utility function U : (0, 1) $\xrightarrow{}$ R is smooth, strictly increasing and strictly concave. 

If we calculate the Arrow Pratt measures of (absolute and relative) risk aversion (RA) defined as $RRA = - \frac{U''(x)}{U'(x)} x$. This is used to determine the range of the parameter of the utility function. 
\begin{itemize}
\item
A common utility function we use in finance is the power utility $U(x) = \frac{x^\gamma}{\gamma}$. We get the $RRA = 1 - \gamma$. If $ 1- \gamma > 0$ then the agent is risk averse. If $1 - \gamma < 0$, we would call her risk seeker. Power utility function with $1 - \gamma> 0$ refers to an investor with RRA that is independent of her level of wealth, which is why it is also called the constant RRA utility function.
\item
We can also consider a logarithmic utility function $U(x) = log (x)$. The RRA value obtained is 1, also a constant.
\item
Another commonly used utility function is the negative exponential utility function $U(x) = -\frac{e^{-\eta x}}{\eta}$. We can estimate the $RRA = \eta x$. This is why this utility function is called the Constant Absolute Relative Risk Aversion (CARA) utility function. For an investor to be risk averse, we would require $\eta > 0.$
\item
If we consider a linear utility function $U(x) = a + bx$, it corresponds to risk neutral investor as RRA = 0.
\end{itemize}

So considering all the different types of utility function, in particular, we specialise to the constant relative risk aversion utility function $U(x) = \frac{x^\gamma}{\gamma}$ for explicit exposition of the derivation of our results. For the objective function, we take the usual time discounted aggregate utility.
Then the objective function which needs to be maximized is subjected to the condition that with more than 95\% probability the lower 5\% quantile of the wealth process should be above some critical value (we denote this by $Q_{0.05}$).

We denote this lower quantile process by $L^{(2)}$, whose equation of motion is given by
\begin{equation}
    d L_{t}^{(2)} = L_{t}^{(2)}(1 - \pi_t) r dt + L_{t}^{(2)} \pi_t d X_2 \nonumber
\end{equation}
\begin{equation}
\textit{or, }    d L_{t}^{(2)} = L_{t}^{(2)}(1 - \pi_t) r dt + L_{t}^{(2)} \pi_tb_2 dt + L_{t}^{(2)} \pi_t \Big(\frac{p_1(1-p_2)}{f_1f_2}dW_t^{(1)} +  \frac{p_2(1-p_2)}{f_2^2} dW_t^{(2)}\Big) \nonumber
\end{equation}

\begin{equation}
\textit{or, }     d L_{t}^{(2)} = L_{t}^{(2)}(r - r\pi_t + b_2\pi_t) dt + L_{t}^{(2)} \pi_t \frac{p_1(1-p_2)}{f_1f_2} dW_t^{(1)} + L_{t}^{(2)} \pi_t \frac{p_2(1-p_2)}{f_2^2} dW_t^{(2)}.\nonumber
\end{equation}
This should be above $Q_{0.05}$ with a high probability. Mathematically this optimization problem with a single constraint can be written as

\begin{equation}\label{eq2}
\begin{aligned}
& \underset{\pi_t}{\text{maximize}}
& & \mathbbm{E} \Bigg(\int_{0}^{T} e^{-\beta t} \frac{L_{t}^{(1)\gamma}}{\gamma} dt \Bigg) \\
& \text{subject to}
& & \frac{1}{T}\int_{0}^{T}\text{Pr} \bigg(L_{t}^{(2)} \geq Q_{0.05}\bigg) dt \geq 0.95
\end{aligned}
\end{equation}

where $\beta > 0$ is the rate of discount over time, $\gamma \in (0, 1)$ is the risk aversion parameter, $p_1 = 0.05$, $p_2 = 0.5$ and since it is continuous we have considered the equality constraint.

 We proceed to solve the above constrained stochastic optimal control problem using stochastic maximum principle following \citeauthor{Ref13}\cite{Ref13}. To this end we introduce appropriate notation.
  Note that \eqref{eq1} represents the state equation and \eqref{eq2} represents the objective utility function and the state constraint.


The constraint can be written as
\[\mathbbm{E} \int_{0}^{T} \frac{1}{T}\quad \mathbbm{I} \bigg(L^{(2)}_t \geq Q_{0.05}\bigg) dt \ge 0.95. \]
Let
\begin{equation}
    J_1 (t, l_1, l_2, \pi) = e^{-\beta t} \frac{l_1^{\gamma}}{\gamma} \nonumber
\end{equation}
\begin{equation}
    J_2 (t, l_1, l_2, \pi) = \frac{1}{T} \mathbbm{I} \bigg\{l_2 \geq Q_{0.05}\bigg\}. \nonumber
\end{equation}
Since $J_2$ is not continuous, we approximate it to a continuous and differentiable function with the help of the Sigmoid function where the choice of parameter $\epsilon$ below needs to be made in such a way that it is very very close to the indicator function,

\[ J_{2}^{\epsilon}(t, l_1, l_2, \pi) = \frac{1}{T}\begin{cases}
      0 & l_2\leq Q_{0.05} - \epsilon \\
      \frac{1}{1 + e^{-\alpha (l_2 - Q_{0.05})}} & Q_{0.05} - \epsilon < l_2 < Q_{0.05} + \epsilon \\
      1 & l_2\geq Q_{0.05} + \epsilon.
   \end{cases}
\]
Now we proceed to solve the (approximated) constrained stochastic optimal control problem with $J_2^{\epsilon}$ replacing $J_2$.
For this  set
\[
b (t, l_1, l_2, \pi) =
\begin{bmatrix}
     l_1(1 - \pi)r \\
    l_2(r - r\pi + b_2\pi)
\end{bmatrix}.
\]

\[
\sigma (t, l_1, l_2, \pi) =
\begin{bmatrix}
     l_1\pi \bigg(\frac{p_1(1-p_1)}{f_1^2} \bigg) & l_1 \pi\bigg( \frac{p_1(1-p_2)}{f_1f_2} \bigg) \\
    l_2 \pi \bigg(\frac{p_1(1-p_2)}{f_1f_2} \bigg) &  l_2 \pi \bigg(\frac{p_2(1-p_2)}{f_2^2}\bigg)
\end{bmatrix}.
\]
Define the Hamiltonian $$ \text{H}^{\epsilon} : [0,T] \times \mathbb{R} \times \mathbb{R} \times \mathbb{R} \times \mathbb{R}^2 \times \mathbb{R}^{2 \times 2}
\times \mathbb{R} \times \mathbb{R} \to \mathbb{R}$$ by
\begin{eqnarray}
    \text{H}^{\epsilon} (t, l_1, l_2, \pi, s, q, \psi^{0}, \psi): &=& -\psi^{0} e^{-\beta t} \frac{l_1^{\gamma}}{\gamma} - \psi J_{2}^{\epsilon}(t, l_1, l_2, \pi) + \begin{bmatrix}s_1 & s_2 \end{bmatrix} \begin{bmatrix}
     l_1(1 - \pi)r \\
    l_2(r - r\pi + b_2\pi)\end{bmatrix} +  \nonumber \\&& tr\Bigg\{\begin{bmatrix} q_1 & q_2 \\q_3 & q_4 \end{bmatrix} \begin{bmatrix}
     l_1\pi \bigg(\frac{p_1(1-p_1)}{f_1^2} \bigg) & l_1 \pi\bigg( \frac{p_1(1-p_2)}{f_1f_2} \bigg) \\
    l_2 \pi \bigg(\frac{p_1(1-p_2)}{f_1f_2} \bigg) &  l_2 \pi \bigg(\frac{p_2(1-p_2)}{f_2^2} \bigg)
\end{bmatrix}\Bigg\}. \nonumber
\end{eqnarray}

First order derivative $J^{'\epsilon}_2(t, l_1, l_2, \pi)$ with respect to $l_2$ is given by

\[J_{2}^{'\epsilon}(t, l_1, l_2, \pi) = \frac{1}{T}\begin{cases}
      0 & l_2 \leq Q_{0.05} - \epsilon \\
      \frac{\alpha e^{-\alpha \big(l_2 - Q_{0.05}\big)}}{\bigg(1 + e^{-\alpha \big(l_2 - Q_{0.05}\big)}\bigg)^2} & Q_{0.05} - \epsilon < l_2 < Q_{0.05} + \epsilon \\
      0 & l_2 \geq Q_{0.05} + \epsilon
   \end{cases}
\]

We can represent this in terms of an indicator function as,
\[J_{2}^{'\epsilon}(t, l_1, l_2, \pi) = \frac{1}{T} \frac{\alpha e^{-\alpha \big(l_2 - Q_{0.05}\big)}}{\bigg(1 + e^{-\alpha \big(l_2 - Q_{0.05}\big)}\bigg)^2} \mathbbm{I} \Bigg(Q_{0.05} - \epsilon < l_2 < Q_{0.05} + \epsilon \Bigg)\]

\[\textit{or, }J_{2}^{'\epsilon}(t, l_1, l_2, \pi) = \frac{1}{T} \alpha  e^{-\alpha \big(l_2 - Q_{0.05}\big)}\bigg(1 + e^{-\alpha \big(l_2 - Q_{0.05}\big)}\bigg)^{-2} \mathbbm{I} \Bigg(Q_{0.05} - \epsilon < l_2 < Q_{0.05} + \epsilon \Bigg).\]

Now following Theorem 6.1 of \citeauthor{Ref13}\cite{Ref13} we state the following result; the proof follows from \citeauthor{Ref13}\cite{Ref13}, pp. 144-153.\\

\textbf{Theorem:}
Let $(\Bar{L_1}(\cdot), \Bar{L_2}(\cdot), \Bar{\pi}(\cdot))$ be  optimal  for the approximated constrained problem.
Then there exists
 $(\psi^{0}, \psi) \in \mathbb{R} \times \mathbb{R}$ satisfying
  $$\psi^0 \geq 0 , \: (\psi^0)^2 + \psi^2 = 1,$$
$$ \psi (z  + \int_{0}^{T} \mathbbm{E} J^{\epsilon}_{2}(t, \bar{L}_1(t), \bar{L}_2(t), \bar{\pi_t})\: dt) \geq 0, \: \: \forall z \in [0.95, 1]$$
 and $\{{\mathcal{F}}_t \}$-adapted solutions
 $$(s(.),q(.)) \in L^2(0, T: \mathbb{R}^2) \times L^2(0, T: \mathbb{R}^2), \: \: (S(.),Q(.)) \in  L^2(0, T: {\mathcal{S}}^2) \times L^2(0, T: {\mathcal{S}}^2) $$
 (where ${\mathcal{S}}^2$ denotes the space of  $2 \times 2$  matrices) of the following adjoint equations:

\begin{equation}\label{eq4}
    \begin{bmatrix}ds_1(t) \\ ds_2(t)\end{bmatrix} = - \begin{bmatrix} \text{H}^{\epsilon}_{l_1}(t, \Bar{L_1}(t), \Bar{L_2}(t), \Bar{\pi_t}, s(t), q(t), \psi^{0}, \psi) \\ \text{H}^{\epsilon}_{l_2}(t, \Bar{L_1}(t), \Bar{L_2}(t), \Bar{\pi_t}, s(t), q(t), \psi^{0}, \psi)\end{bmatrix} dt  + q(t) \begin{bmatrix} dW_t^{(1)} \\ dW_t^{(2)}\end{bmatrix}
\end{equation}
with boundary condition
\begin{equation}    \begin{bmatrix}s_1(T)\\s_2(T)\end{bmatrix} = 0, \nonumber
\end{equation}
and
\begin{eqnarray}\label{adjoint1}
dS(t) &=& -\Bigg\{b_x(t, \Bar{L_1}(t), \Bar{L_2}(t), \Bar{\pi_t})^{T}S(t) + S(t)b_x(t, \Bar{L_1}(t), \Bar{L_2}(t),  \Bar{\pi_t}) + \sum_{j=1}^{2}\sigma^{j}_{x}(t, \Bar{L_1}(t), \Bar{L_2}
(t), \Bar{\pi_t})^{T}S(t)\sigma^{j}_{x}(t, \Bar{L_1}(t), \Bar{L_2}(t),  \Bar{\pi_t}) \nonumber \\&&  + \sum_{j=1}^{2}\sigma^{j}_{x}(t, \Bar{L_1}(t), \Bar{L_2}(t),  \Bar{\pi_t})^{T}Q_{j}(t) + Q_{j}(t)\sigma^{j}_{x}(t, \Bar{L_1}(t), \Bar{L_2}(t), \Bar{\pi_t}) + H^{\epsilon}_{xx}(t, \Bar{L_1}(t), \Bar{L_2}(t), \Bar{\pi_t}, s(t), q(t), \psi, \psi^{0})\bigg\} + \nonumber \\&& \sum_{j=1}^{2}Q_{j}(t)dW^{j}(t)
\end{eqnarray}
where $x$ is the vector $(l_1,l_2)$, with boundary condition
\begin{equation}    \begin{bmatrix}S_1(T)\\S_2(T)\end{bmatrix} = 0, \nonumber
 \end{equation}
 such that for
 $\mathcal{H}^{\epsilon} (t, l_1, l_2, \pi)$ defined by
\begin{eqnarray}
   \mathcal{H}^{\epsilon} (t, l_1, l_2, \pi) &=& \text{H}^{\epsilon} (t, l_1, l_2, \pi, s, q, \psi, \psi^{0}) - \frac{1}{2} tr\Big\{\sigma(t, \Bar{l_1}, \Bar{l_2}, \Bar{\pi})^{T} S(t) \sigma(t, \Bar{l_1}, \Bar{l_2}, \Bar{\pi})\Big\}  + \nonumber \\&& \frac{1}{2} tr\Big\{\big[\sigma(t, l_1, l_2, \pi) - \sigma(t, \Bar{l_1}, \Bar{l_2}, \Bar{\pi})\big]^{T} S(t) \big[\sigma(t, l_1, l_2, \pi) - \sigma(t, \Bar{l_1}, \Bar{l_2}, \Bar{\pi})\big]\Big\}\nonumber
\end{eqnarray}
satisfies
\begin{equation}\label{condition}
   \mathcal{H}^{\epsilon} (t, \Bar{L_1}(t), \Bar{L_2}(t), \Bar{\pi_t}) = \sup_{\pi \in \mathbb{R}} \mathcal{H}^{\epsilon} (t, \Bar{L_1}(t), \Bar{L_2}(t), \pi_t).
\end{equation} \qed

\vspace{3mm}

The above theorem gives us a complete solution to the portfolio optimisation problem that we set out to do in an implicit manner. Our next step is to solve the above adjoint equation (which are backward stochastic differential equations) and use the Hamiltonian maximization to find an optimal portfolio explicitly.

For simplification let us denote $k_1 = J_{2}^{'\epsilon}(t, l_1, l_2, \pi)$,
\begin{equation}
    k_1 = \frac{1}{T} \alpha  e^{-\alpha \big(l_2 - Q_{0.05}\big)}\bigg(1 + e^{-\alpha \big(l_2 - Q_{0.05}\big)}\bigg)^{-2} \mathbbm{I} \Bigg(Q_{0.05} - \epsilon < l_2 < Q_{0.05} + \epsilon \Bigg). \nonumber
\end{equation}

 First consider the adjoint equation  \eqref{eq4}, i.e.,
\begin{eqnarray}
\begin{bmatrix} ds_1(t) \\ ds_2(t) \end{bmatrix} &=& \begin{bmatrix}\psi^{0} e^{-\beta t} \Bar{L_1}(t)^{\gamma - 1} -  s_1(t) (1 -  \Bar{\pi_t}) r  \\ k_1\psi - s_2(t)(r - r\Bar{\pi}_t + b_2 \Bar{\pi_t})\end{bmatrix} dt + q(t)\begin{bmatrix}dW_t^{(1)} \\ dW_t^{(2)}\end{bmatrix} \nonumber
\end{eqnarray}
\begin{equation}
\textit{and }    \begin{bmatrix}s_1(T)\\s_2(T)\end{bmatrix} = 0. \nonumber
\end{equation}

Choose q(t) = 0 which reduces the SDE in to random ODEs in $s_1(t)$ and $s_2(t)$
\begin{eqnarray}
    ds_1(t) &=& \big\{\psi^{0} e^{-\beta t} \Bar{L_1}(t)^{\gamma - 1}  - s_1(t) (1 -  \Bar{\pi_t}) r \big\} dt \nonumber
\end{eqnarray}
\begin{eqnarray}
\textit{or }    \frac{ds_1(t)}{dt} + s_1(t) (1 -  \Bar{\pi_t}) r &=& \psi^{0} e^{-\beta t} \Bar{L_1}(t)^{\gamma - 1}\nonumber
\end{eqnarray}
Integrating, we have
\begin{eqnarray}
    s_1(t) = \frac{\psi^{0} e^{-\beta t} \Bar{L_1}(t)^{\gamma - 1}}{(1 -  \Bar{\pi_t}) r-\beta} + c_1 e^{-(1 -  \Bar{\pi_t}) r t}. \nonumber
\end{eqnarray}
Putting the terminal constraint we get
\begin{eqnarray}
    s_1(t) = \frac{\psi^{0} e^{-\beta t} \Bar{L_1}(t)^{\gamma - 1}}{(1 - \Bar{\pi_t}) r-\beta} -  \frac{\psi^{0} e^{-\beta T} \Bar{L_1}(t)^{\gamma - 1}}{(1 - \Bar{\pi_t}) r-\beta} e^{-(1 - \Bar{\pi}_t) r (t - T)}. \nonumber
\end{eqnarray}
Again,
\begin{eqnarray}
    ds_2(t) &=& \big\{\psi k_1 - s_2(t) (r -  \Bar{\pi_t} r + b_2 \Bar{\pi_t}) \big\} dt \nonumber
\end{eqnarray}
\begin{eqnarray}
\textit{or }    \frac{ds_2(t)}{dt} + s_2(t) (r -  \Bar{\pi_t} r + b_2 \Bar{\pi_t}) &=& \psi k_1 \nonumber
\end{eqnarray}
Integrating, we have
\begin{eqnarray}
    s_2(t) = \frac{\psi k_1}{(r -  \Bar{\pi_t} r + b_2 \Bar{\pi_t})} + c_2 e^{-(r -  \Bar{\pi_t} r + b_2 \Bar{\pi_t}) t}. \nonumber
\end{eqnarray}
Putting the terminal constraint we get
\begin{eqnarray}
    s_2(t) = \frac{\psi k_1}{(r -  \Bar{\pi_t} r + b_2 \Bar{\pi_t})} -  \frac{\psi k_1}{(r -  \Bar{\pi_t} r + b_2 \Bar{\pi_t})} e^{-(r -  \Bar{\pi_t} r + b_2 \Bar{\pi_t}) (t - T)}. \nonumber
\end{eqnarray}
Therefore $s(t)$ is given by
\begin{equation}
    s(t) = \begin{bmatrix}
         s_1(t) \\ s_2(t)
    \end{bmatrix} = \begin{bmatrix}
     \frac{\psi^{0} e^{-\beta t} \Bar{L_1}(t)^{\gamma - 1}}{(1 -  \Bar{\pi_t}) r-\beta} -  \frac{\psi^{0} e^{-\beta T} \Bar{L_1}(t)^{\gamma - 1}}{(1 - \Bar{\pi_t}) r-\beta} e^{-(1 - \Bar{\pi_t}) r (t - T)} \\
     \frac{\psi k_1}{(r - \Bar{\pi_t} r + b_2 \Bar{\pi_t})} -  \frac{\psi k_1}{(r -  \Bar{\pi_t} r + b_2 \Bar{\pi_t})} e^{-(r -  \Bar{\pi_t} r + b_2 \Bar{\pi_t}) (t - T)} \nonumber
    \end{bmatrix}.
\end{equation}
Thus $(s(t), 0)$ is the unique solution of the adjoint equation \eqref{eq4}.\\

Next, the second order derivative $J^{''\epsilon}_{2}(t, l_1, l_2, \pi)$ with respect to $l_2$ is given by

\[J^{''\epsilon}_{2}(t, l_1, l_2, \pi) = \frac{1}{T}\begin{cases}     0 & l_2 \leq Q_{0.05} - \epsilon \\
       \frac{\alpha^2 e^{-\alpha(l_2 - Q_{0.05})}}{\bigg(1 + e^{-\alpha \big(l_2 - Q_{0.05}\big)}\bigg)^2} - \frac{2\bigg(\alpha e^{-\alpha \big(l_2 - Q_{0.05}\big)}\bigg)^2}{\bigg(1 + e^{-\alpha \big(l_2 - Q_{0.05}\big)}\bigg)^3} & Q_{0.05} - \epsilon < l_2 < Q_{0.05} + \epsilon \\
      0 & l_2 \geq Q_{0.05} + \epsilon
   \end{cases}
\]

We can represent $k_2 = J^{''\epsilon}_{2}(t, l_1, l_2, \pi)$, so writing $k_2$ in terms of indicator function as
\begin{eqnarray}
  k_2 &=& \frac{1}{T}\Bigg[\frac{\alpha^2 e^{-\alpha (l_2 - Q_{0.05})}}{\bigg(1 + e^{-\alpha \big(l_2 - Q_{0.05}\big)}\bigg)^2} - \frac{2\bigg(\alpha  e^{-\alpha \big(l_2 - Q_{0.05}\big)}\bigg)^2}{\bigg(1 + e^{-\alpha \big(l_2 - Q_{0.05}\big)}\bigg)^3}\Bigg] \mathbbm{I}\bigg(Q_{0.05} - \epsilon < l_2 < Q_{0.05} + \epsilon\bigg)  \nonumber \\
\textit{or } k_2
  &=& \frac{1}{T}\Bigg[\alpha k_1 \frac{1 - e^{-\alpha \big(l_2 - Q_{0.05}\big)}}{1 + e^{-\alpha \big(l_2 - Q_{0.05}\big)}}\Bigg] \mathbbm{I}\bigg(Q_{0.05} - \epsilon < l_2 < Q_{0.05} + \epsilon\bigg)\nonumber
\end{eqnarray}

Consider the second adjoint equation  \eqref{adjoint1},
\begin{eqnarray}
dS(t) &=& -\Bigg\{b_x(t, \Bar{L_1}(t), \Bar{L_2}(t),  \Bar{\pi_t})^{T}S(t) + S(t)b_x(t, \Bar{L_1}(t), \Bar{L_2}(t),  \Bar{\pi_t}) + \sum_{j=1}^{2}\sigma^{j}_{x}(t, \Bar{L_1}(t), \Bar{L_2}
(t), \Bar{\pi_t})^{T}S(t)\sigma^{j}_{x}(t, \Bar{L_1}(t), \Bar{L_2}(t),  \Bar{\pi_t}) \nonumber \\&&  + \sum_{j=1}^{2}\sigma^{j}_{x}(t, \Bar{L_1}(t), \Bar{L_2}(t), \Bar{\pi_t})^{T}Q_{j}(t) + Q_{j}(t)\sigma^{j}_{x}(t, \Bar{L_1}(t), \Bar{L_2}(t),  \Bar{\pi_t}) + H^{\epsilon}_{xx}(t, \Bar{L_1}(t), \Bar{L_2}(t),  \Bar{\pi_t}, s(t), q(t), \psi, \psi^{0})\bigg\} + \nonumber \\&& \sum_{j=1}^{2}Q_{j}(t)dW^{j}(t) \nonumber
\end{eqnarray}
Again we choose $Q(t)=0$ to reduce the SDE in to random ODEs given by
\begin{eqnarray}
\begin{bmatrix}
   dS_{11}(t) & dS_{12}(t) \\ dS_{21}(t) & dS_{22}(t)
\end{bmatrix} &=& -\Bigg\{\begin{bmatrix}
     (1 -  \Bar{\pi_t}) r & 0 \\ 0 & (r - r \Bar{\pi_t} + b_2 \Bar{\pi_t})
\end{bmatrix}\begin{bmatrix}
   S_{11}(t) & S_{12}(t) \\ S_{21}(t) & S_{22}(t)
\end{bmatrix} +
\begin{bmatrix}
   S_{11}(t) & S_{12}(t) \\ S_{21}(t) & S_{22}(t)
   \end{bmatrix}\begin{bmatrix}
     (1 - \Bar{\pi_t}) r & 0 \\ 0 & (r - r \Bar{\pi_t} + b_2 \Bar{\pi_t})
\end{bmatrix}
+ \nonumber \\&&
\begin{bmatrix}
    \Bar{\pi_t}\bigg(\frac{p_1(1-p_1)}{f_1^2} \bigg) & 0 \\0 & \Bar{\pi_t}\bigg(\frac{p_1(1-p_2)}{f_1f_2} \bigg)
\end{bmatrix}\begin{bmatrix}
   S_{11}(t) & S_{12}(t) \\ S_{21}(t) & S_{22}(t)
   \end{bmatrix} \nonumber \\&& \begin{bmatrix}
   \Bar{\pi_t}\bigg(\frac{p_1(1-p_1)}{f_1^2} \bigg) & 0 \\0 & \Bar{\pi_t}\bigg(\frac{p_1(1-p_2)}{f_1f_2} \bigg)
\end{bmatrix} + \nonumber \\&&
\begin{bmatrix}
   \Bar{\pi_t}\bigg(\frac{p_1(1-p_2)}{f_1f_2}\bigg) & 0 \\0 &  \Bar{\pi_t}\bigg(\frac{p_2(1-p_2)}{f_2^2}\bigg)
\end{bmatrix}\begin{bmatrix}
   S_{11}(t) & S_{12}(t) \\ S_{21}(t) & S_{22}(t)
   \end{bmatrix} \nonumber \\&& \begin{bmatrix}
    \Bar{\pi_t}\bigg(\frac{p_1(1-p_2)}{f_1f_2}\bigg) & 0 \\0 &  \Bar{\pi_t}\bigg(\frac{p_2(1-p_2)}{f_2^2}\bigg)
\end{bmatrix} + \nonumber \\&&
\begin{bmatrix} \text{H}^{\epsilon}_{l_1l_1}(t, \Bar{L_1}(t), \Bar{L_2}(t),  \Bar{\pi_t}, s(t), q(t), \psi^{0}, \psi) & \text{H}^{\epsilon}_{l_1l_2}(t, \Bar{L_1}(t), \Bar{L_2}(t),  \Bar{\pi_t}, s(t), q(t), \psi^{0}, \psi) \\ \text{H}^{\epsilon}_{l_2l_1}(t, \Bar{L_1}(t), \Bar{L_2}(t), \Bar{\pi_t}, s(t), q(t), \psi^{0}, \psi) & \text{H}^{\epsilon}_{l_2l_2}(t, \Bar{L_1}(t), \Bar{L_2}(t),  \Bar{\pi_t}, s(t), q(t), \psi^{0}, \psi)\end{bmatrix} \Bigg\} dt \nonumber
\end{eqnarray}

\begin{eqnarray}
\begin{bmatrix}
 dS_{11}(t) & dS_{12}(t) \\ dS_{21}(t) & dS_{22}(t)
\end{bmatrix} &=& -\Bigg\{2\begin{bmatrix}
     (1 - \Bar{\pi_t}) r S_{11}(t) & (1 -  \Bar{\pi_t}) r S_{12}(t) \\ (r - r \Bar{\pi_t} + b_2 \Bar{\pi_t})S_{21}(t) & (r - r \Bar{\pi_t} + b_2 \Bar{\pi_t}) S_{22}(t)
\end{bmatrix} +
 \nonumber \\&&
\begin{bmatrix}
    \Bar{\pi_t}^2\bigg(\frac{p_1(1-p_1)}{f_1^2} \bigg)^2S_{11}(t) & A_{12}S_{12}(t) \\ A_{21}S_{21}(t) &  \Bar{\pi_t}^2\bigg(\frac{p_1(1-p_2)}{f_1f_2}\bigg)^2S_{22}(t)
\end{bmatrix} + \nonumber \\&&
\begin{bmatrix}
    \Bar{\pi_t}^2\bigg(\frac{p_1(1-p_2)}{f_1f_2}\bigg)^2S_{11}(t) & B_{12}S_{12}(t) \\ B_{21}S_{21}(t) &  \Bar{\pi_t}^2\bigg(\frac{p_2(1-p_2)}{f_2^2}\bigg)^2S_{22}(t)
\end{bmatrix} + \nonumber \\&&
\begin{bmatrix} \psi^{0} e^{-\beta t} (\gamma - 1)\Bar{L_1}(t)^{\gamma - 2}  & 0 \\ 0 & \psi k_2\end{bmatrix} \Bigg\} dt \nonumber
\end{eqnarray}
where
\begin{equation}
A_{12} = A_{21} =  \Bar{\pi_t}\bigg(\frac{p_1(1-p_1)}{f_1^2} \bigg) \Bar{\pi_t}\bigg(\frac{p_1(1-p_2)}{f_1f_2} \bigg) \nonumber
\end{equation}
\begin{equation}
B_{12} = B_{21} = \Bar{\pi_t}\bigg(\frac{p_1(1-p_1)}{f_1f_2}\bigg) \Bar{\pi_t}\bigg(\frac{p_2(1-p_2)}{f_2^2}\bigg) \nonumber
\end{equation}
Solving this for each element, we obtain

\begin{eqnarray}
dS_{11}(t) &=& \bigg\{- 2(1 -  \Bar{\pi_t}) r S_{11}(t) -  \Bar{\pi_t}^2\bigg(\frac{p_1(1-p_1)}{f_1^2} \bigg)^2S_{11}(t) -  \Bar{\pi_t}^2\bigg(\frac{p_1(1-p_2)}{f_1f_2}\bigg)^2S_{11}(t) -  \psi^{0} e^{-\beta t} (\gamma - 1)\Bar{L_1}(t)^{\gamma - 2} \bigg\} dt \nonumber
\end{eqnarray}
\begin{equation}
\textit{and } S_{11}(T) = 0. \nonumber
\end{equation}
\begin{eqnarray}
\textit{or }   \frac{dS_{11}(t)}{dt} &=& \bigg\{- 2(1 -  \Bar{\pi_t}) r  -  \Bar{\pi_t}^2\bigg(\frac{p_1(1-p_1)}{f_1^2} \bigg)^2 -  \Bar{\pi_t}^2\bigg(\frac{p_1(1-p_2)}{f_1f_2}\bigg)^2\bigg\} S_{11}(t) -  \psi^{0} e^{-\beta t} (\gamma - 1)\Bar{L_1}(t)^{\gamma - 2} \nonumber
\end{eqnarray}

\begin{eqnarray}
\textit{or }    \frac{dS_{11}(t)}{dt} + \bigg\{ 2(1 -  \Bar{\pi_t}) r  + \Bar{\pi_t}^2\bigg(\frac{p_1(1-p_1)}{f_1^2} \bigg)^2 +  \Bar{\pi_t}^2\bigg(\frac{p_1(1-p_2)}{f_1f_2}\bigg)^2\bigg\} S_{11}(t) =  - \psi^{0} e^{-\beta t} (\gamma - 1)\Bar{L_t}^{(1)\gamma - 2}. \nonumber
\end{eqnarray}

Integrating and using the terminal conditions we get,

\begin{eqnarray}
    S_{11}(t) &=&\frac{-\psi^{0} e^{-\beta t} (\gamma - 1)\Bar{L_1}(t)^{\gamma - 2}}{\Bigg(2(1 -  \Bar{\pi_t}) r  +  \Bar{\pi_t}^2\bigg(\frac{p_1(1-p_1)}{f_1^2} \bigg)^2 +  \Bar{\pi_t}^2\bigg(\frac{p_1(1-p_2)}{f_1f_2}\bigg)^2 - \beta \Bigg)} +  \frac{\psi^{0} e^{-\beta T} (\gamma - 1)\Bar{L_1}(t)^{\gamma - 2}}{\Bigg(2(1 -  \Bar{\pi_t}) r  +  \Bar{\pi_t}^2\bigg(\frac{p_1(1-p_1)}{f_1^2} \bigg)^2 +  \Bar{\pi_t}^2\bigg(\frac{p_1(1-p_2)}{f_1f_2}\bigg)^2 - \beta \Bigg)} \times \nonumber \\&&
    e^{-\Big(2(1 -  \Bar{\pi_t}) r  +  \Bar{\pi_t}^2\bigg(\frac{p_1(1-p_1)}{f_1^2} \bigg)^2 +  \Bar{\pi_t}^2\bigg(\frac{p_1(1-p_2)}{f_1f_2}\bigg)^2 (t-T)\Big)} \nonumber
\end{eqnarray}
Again,
\begin{eqnarray}
dS_{12}(t) &=& \bigg\{- 2(1 -  \Bar{\pi_t}) r S_{12}(t) -  \Bar{\pi_t}\bigg(\frac{p_1(1-p_1)}{f_1^2} \bigg) \Bar{\pi_t}\bigg(\frac{p_1(1-p_2)}{f_1f_2} \bigg)S_{12}(t) -   \Bar{\pi_t}\bigg(\frac{p_1(1-p_1)}{f_1f_2}\bigg) \Bar{\pi_t}\bigg(\frac{p_2(1-p_2)}{f_2^2}\bigg)S_{12}(t) \bigg\} dt \nonumber
\end{eqnarray}
\begin{equation}
\textit{and } S_{12}(T) = 0. \nonumber
\end{equation}
\begin{eqnarray}
\textit{or }   \frac{dS_{12}(t)}{S_{12}(t)} &=& \bigg\{- 2(1 -  \Bar{\pi_t}) r  -  \Bar{\pi_t}\bigg(\frac{p_1(1-p_1)}{f_1^2} \bigg) \Bar{\pi_t}\bigg(\frac{p_1(1-p_2)}{f_1f_2} + \bigg) -  \Bar{\pi_t}\bigg(\frac{p_1(1-p_1)}{f_1f_2}\bigg) \Bar{\pi_t}\bigg(\frac{p_2(1-p_2)}{f_2^2}\bigg) \bigg\} dt \nonumber
\end{eqnarray}
Integrating and using the terminal conditions we get,

\begin{eqnarray}
    S_{12}(t) &=&e^{-\bigg\{2(1 - \Bar{\pi_t}) r  +  \Bar{\pi_t}\bigg(\frac{p_1(1-p_1)}{f_1^2} \bigg) \Bar{\pi_t}\bigg(\frac{p_1(1-p_2)}{f_1f_2}\bigg) +  \Bar{\pi_t}\bigg(\frac{p_1(1-p_1)}{f_1f_2}\bigg) \Bar{\pi_t}\bigg(\frac{p_2(1-p_2)}{f_2^2}\bigg) \bigg\} t} \nonumber
\end{eqnarray}
The other two elements can be similarly found as,
\begin{eqnarray}
    S_{21}(t) &=&e^{-\bigg\{2(r - r \Bar{\pi_t} + b_2 \Bar{\pi_t})  +  \Bar{\pi_t}\bigg(\frac{p_1(1-p_1)}{f_1^2} \bigg) \Bar{\pi_t}\bigg(\frac{p_1(1-p_2)}{f_1f_2} \bigg) +  \Bar{\pi_t}\bigg(\frac{p_1(1-p_1)}{f_1f_2}\bigg) \Bar{\pi_t}\bigg(\frac{p_2(1-p_2)}{f_2^2}\bigg) \bigg\} t} \nonumber
\end{eqnarray}
\begin{eqnarray}
\textit{and }    S_{22}(t) &=&\frac{-\psi k_2}{\Bigg(2(r - r \Bar{\pi_t} + b_2 \Bar{\pi_t})  +  \Bar{\pi_t}^2\bigg(\frac{p_2(1-p_2)}{f_2^2} \bigg)^2 +  \Bar{\pi_t}^2\bigg(\frac{p_2(1-p_2)}{f_2^2}\bigg)^2 - \beta \Bigg)} + \nonumber \\&& \frac{\psi k_2}{\Bigg(2(1 -  \Bar{\pi_t}) r  +  \Bar{\pi_t}^2\bigg(\frac{p_2(1-p_2)}{f_2^2} \bigg)^2 +  \Bar{\pi_t}^2\bigg(\frac{p_2(1-p_2)}{f_2^2}\bigg)^2 - \beta \Bigg)} \times \nonumber \\&&
    e^{-\Big(2(1 -  \Bar{\pi_t}) r  +  \Bar{\pi_t}^2\bigg(\frac{p_2(1-p_2)}{f_2^2}\bigg)^2 +  \Bar{\pi_t}^2\bigg(\frac{p_2(1-p_2)}{f_2^2}\bigg)^2 (t-T)\Big)} \nonumber
\end{eqnarray}

Thus $(S(t), 0)$ is the unique solution of the adjoint equation \eqref{adjoint1}.\\

Therefore the unique adapted solution to the  adjoint equations \eqref{eq4} \& \eqref{adjoint1} are given by the following pairs:

\begin{eqnarray}
\Bigg(\begin{bmatrix}s_1(t) \\s_2(t) \end{bmatrix}, 0\Bigg) & \textit{and } &
\Bigg(\begin{bmatrix}S_{11}(t) & S_{12}(t) \\S_{21}(t) & S_{22}(t) \end{bmatrix}, 0\Bigg). \nonumber
\end{eqnarray}

Now in the next step to get the optimal points satisfying \eqref{condition}, we can write it as,
\begin{equation}
   \max_{\pi \in \Pi} \mathcal{H}^{\epsilon} (t, \Bar{L_1}(t), \Bar{L_2}(t), \pi_t) : \frac{d}{d \pi}\big\{\mathcal{H}^{\epsilon} (t, \Bar{L_1}(t), \Bar{L_2}(t), \pi_t)\big\} = 0 \nonumber
\end{equation}
\begin{eqnarray}
  \frac{d}{d \pi}\big\{\mathcal{H}^{\epsilon} (t, \Bar{L_1}(t), \Bar{L_2}(t), \pi_t)\big\} &=& \frac{d}{d \pi}\Bigg( \text{H}^{\epsilon} (t, \Bar{L_1}(t), \Bar{L_2}(t), \pi_t, s(t), q(t)) - \frac{1}{2} tr\Big\{\sigma(t, \Bar{L_1}(t), \Bar{L_2}(t), \Bar{\pi_t})^{T} S(t) \sigma(t, \Bar{L_1}(t), \Bar{L_2}(t), \Bar{\pi_t})\Big\}  + \nonumber \\&& \frac{1}{2} tr\Big\{\big[\sigma(t, \Bar{L_1}(t), \Bar{L_2}(t), \pi_t) - \sigma(t, \Bar{L_1}(t), \Bar{L_2}(t), \Bar{\pi_t})\big]^{T} S(t) \nonumber \\&& \big[\sigma(t, \Bar{L_1}(t), \Bar{L_2}(t), \pi_t) - \sigma(t, \Bar{L_1}(t), \Bar{L_2}(t), \Bar{\pi_t})\big]\Big\}\Bigg)  \nonumber
\end{eqnarray}
Or,
\begin{eqnarray}
\frac{d}{d \pi}\big\{\mathcal{H}^{\epsilon} (t, \Bar{L_1}(t), \Bar{L_2}(t), \pi_t)\big\} & =& \frac{d}{d \pi}\Bigg( \text{H}^{\epsilon} (t, \Bar{L_1}(t), \Bar{L_2}(t), \pi_t, s(t), q(t))  + \frac{1}{2} tr\Big\{\big[\sigma(t, \Bar{L_1}(t), \Bar{L_2}(t), \pi_t) - \sigma(t, \Bar{L_1}(t), \Bar{L_2}(t), \Bar{\pi_t})\big]^{T} S(t) \nonumber \\&&  \big[\sigma(t, \Bar{L_1}(t), \Bar{L_2}(t), \pi_t) - \sigma(t, \Bar{L_1}(t), \Bar{L_2}(t), \Bar{\pi_t})\big]\Big\}\Bigg)  \nonumber
\end{eqnarray}
Simplifying we get,
\begin{eqnarray}
 \frac{d}{d \pi}\Big(\frac{1}{2} tr\Big\{\big[\sigma(t, \Bar{L_1}(t), \Bar{L_2}(t), \pi_t) - \sigma(t, \Bar{L_1}(t), \Bar{L_2}(t), \Bar{\pi_t})\big]^{T} S(t)  \big[\sigma(t, \Bar{L_1}(t), \Bar{L_2}(t), \pi_t) - \sigma(t, \Bar{L_1}(t), \Bar{L_2}(t), \Bar{\pi_t})\big]\Big\}\Big) =\nonumber \\  \frac{d}{d \pi}\Big( (\pi_t - \Bar{\pi_t})^2 \times constant \Big) =  0   \nonumber
\end{eqnarray}
and,
\begin{eqnarray}
 \frac{d}{d \pi}\Big(\text{H}^{\epsilon} (t, \Bar{L_1}(t), \Bar{L_2}(t), \pi_t, s(t), q(t))\Big) = -s_1(t)\Bar{L_1}(t)r + s_2(t)\Bar{L_2}(t)(b_2 - r).  \nonumber
\end{eqnarray}
Therefore,
\begin{eqnarray}
  \frac{d}{d \pi}\big\{\mathcal{H}^{\epsilon} (t, \Bar{L_1}(t), \Bar{L_2}(t), \pi_t)\big\} &=&
  s_2(t)\Bar{L_2}(t)(b_2 - r) - s_1(t)\Bar{L_1}(t)r. \nonumber
\end{eqnarray}
Expanding, we rewrite
\begin{eqnarray}
\frac{d}{d \pi}\big\{\mathcal{H}^{\epsilon} (t, \Bar{L_1}(t), \Bar{L_2}(t), \pi_t)\big\} &=&  \Bigg(\frac{\psi k_1}{(r - \Bar{\pi_t} r + b_2\Bar{\pi_t})} -  \frac{\psi k_1}{(r - \Bar{\pi_t} r + b_2\Bar{\pi_t})} e^{-(r - \Bar{\pi_t} r + b_2\Bar{\pi_t}) (t - T)}\Bigg)\Bar{L_2}(t)(b_2 - r) - \nonumber \\&& \Bigg(\frac{\psi^{0} e^{-\beta t} \Bar{L_1}(t)^{\gamma - 1}}{(1 - \Bar{\pi_t}) r-\beta} -  \frac{\psi^{0} e^{-\beta T} \Bar{L_1}(t)^{\gamma - 1}}{(1 - \Bar{\pi_t}) r-\beta} e^{-(1 - \Bar{\pi_t}) r (t - T)}\Bigg)\Bar{L_1}(t)r =  0. \nonumber
\end{eqnarray}
\begin{eqnarray}
\textit{Or, }    \Bigg(\frac{\psi k_1}{(r - \Bar{\pi_t} r + b_2\Bar{\pi_t})} -  \frac{\psi k_1}{(r - \Bar{\pi_t} r + b_2\Bar{\pi_t})} e^{-(r -\Bar{\pi_t} r + b_2\Bar{\pi_t}) (t - T)}\Bigg)\Bar{L_2}(t)(b_2 - r) =  \Bigg(\frac{\psi^{0} e^{-\beta t} \Bar{L_1}(t)^{\gamma}}{(1 - \Bar{\pi_t}) r-\beta} -  \frac{\psi^{0} e^{-\beta T} \Bar{L_1}(t)^{\gamma}}{(1 - \Bar{\pi_t}) r-\beta} e^{-(1 - \Bar{\pi_t}) r (t - T)}\Bigg)r  \nonumber
\end{eqnarray}
\begin{eqnarray}
 \textit{or, }   \Bigg(\frac{\psi k_1}{(r - \Bar{\pi_t} r + b_2\Bar{\pi_t})} \Big\{1 - e^{-(r - \Bar{\pi_t} r + b_2\Bar{\pi_t}) (t - T)}\Big\}\Bigg)\Bar{L_2}(t)(b_2 - r) -  \Bigg(\frac{\psi^{0} \Bar{L_1}(t)^{\gamma}}{(1 - \Bar{\pi_t}) r-\beta} \Big\{e^{-\beta t} -   e^{-\{(1 - \Bar{\pi_t}) r (t - T)  + \beta T\} }\Big\}\Bigg)r = 0 \nonumber
\end{eqnarray}
\begin{eqnarray}
 \textit{or, }    \Bigg(\frac{\psi k_1 \Bar{L_2}(t)(b_2 - r)}{(r - \Bar{\pi_t} r + b_2\Bar{\pi_t})} \Big\{1 - e^{-(r -\Bar{\pi_t} r + b_2\Bar{\pi_t}) (t - T)}\Big\}\Bigg) =  \Bigg(\frac{\psi^{0} r \Bar{L_1}(t)^{\gamma}e^{-\beta t}}{(1 - \Bar{\pi_t}) r-\beta} \Big\{1 -   e^{-\{(1 - \Bar{\pi_t}) r (t - T)  - \beta (t - T)\} }\Big\}\Bigg) \nonumber
\end{eqnarray}
\begin{eqnarray}\label{final}
  \textit{or, }   \frac{\frac{\psi k_1 \Bar{L_2}(t)(b_2 - r)}{(r - \Bar{\pi_t} r + b_2\Bar{\pi_t})}}{\frac{\psi^{0} r \Bar{L_1}(t)^{\gamma}e^{-\beta t}}{(1 - \Bar{\pi_t}) r-\beta}} = \frac{1 - e^{-(r - \Bar{\pi_t} r + b_2\Bar{\pi_t}) (t - T)}}{1 -   e^{-\big((1 -\Bar{\pi_t}) r - \beta\big) (t - T)}} r
\end{eqnarray}

Finally, the optimal portfolio strategy can be obtained by numerically solving  \eqref{final}. We will do this for a real data in the next section where the optimal portfolio will be evaluated for alternative values of the relevant parameters of the equation.

\noindent{\bf Remark.} By Ekeland's thorem \cite{Clarke}, pp. 265-268, it follows that the above solution is an approximate solution of the original problem.


\section{Numerical Illustration}

For the numerical illustration, data used are daily closing price of the stock “Entergy Corporation” in the time range 31st August, 2009 till 30th August, 2013 \citeauthor{Quantopian} \cite{Quantopian}. The return is calculated for this data using the following expression, 
\begin{equation}
    return = \frac{(Price_{t} - Price_{t-1})}{Price_{t-1})}. \nonumber
\end{equation}

We are going to make an approximation of \eqref{final} to get an explicit expression of the optimal strategy.
Since
\begin{eqnarray}
    1 - e^{-(r - \Bar{\pi_t} r + b_2\Bar{\pi_t}) (t - T)} \approx 1 -   e^{-\big((1 - \Bar{\pi_t}) r - \beta\big) (t - T)}, \nonumber
\end{eqnarray}
We can simplify as follows
\begin{eqnarray}
    \frac{\frac{\psi k_1 \Bar{L_2}(t)(b_2 - r)}{(r - \Bar{\pi_t} r + b_2\Bar{\pi_t})}}{\frac{\psi^{0} r \Bar{L_1}(t)^{\gamma}e^{-\beta t}}{(1 - \Bar{\pi_t}) r-\beta}} = \frac{1 - e^{-(r - \Bar{\pi_t} r + b_2\Bar{\pi_t}) (t - T)}}{1 -   e^{-\big((1 - \Bar{\pi_t}) r - \beta\big) (t - T)}} \approx 1\nonumber
\end{eqnarray}
\begin{eqnarray}
 \textit{or, }   \frac{\psi k_1 \Bar{L_2}(t)(b_2 - r)}{(r - \Bar{\pi_t} r + b_2\Bar{\pi_t})} = \frac{\psi^{0} r \Bar{L_1}(t)^{\gamma}e^{-\beta t}}{(1 - \Bar{\pi_t}) r-\beta}. \nonumber
\end{eqnarray}
Thus the optimal portfolio strategy is given by
\begin{eqnarray}\label{approx}
     \Bar{\pi_t} = \frac{\big(e^{-\beta t}\psi^0 \Bar{L_1}(t)^{\gamma} r^2\big) - \big(\psi k_1 \Bar{L_2}(t)(b_2 - r)(r - \beta)\big)}{\big(e^{-\beta t}\psi^0 \Bar{L_1}(t)^{\gamma} - \psi k_1 \Bar{L_2}(t)\big)(b_2 - r)r}.
\end{eqnarray}

Assuming that the distribution function is not known for the return of the stock prices we try to fit a distribution using kernel density estimator (KDE). It is a non parametric way to estimate the probability density function of a random variable. KDE is a fundamental data smoothing problem based on the finite data sample we choose. If we have $(x_1, x_2.....x_n)$ as the independent univariate samples coming from an unknown distribution f then we can write 
\begin{equation}
    \hat{f_h (x)} = \frac{1}{nh}\sum_{i=1}^{n} K \big(\frac{x-x_i}{h}\big) \nonumber
\end{equation}
where K is the kernel which is a non-negative function and h > 0 is a smoothing parameter called the bandwidth \citeauthor{Ref11}\cite{Ref11}. Using the KDE function we get the following estimate of the parameters:
\begin{center}
\begin{tabular}{c c l l}
     kernel density estimate &=& log-quadratic fitting \\  bandwidth (bw) &=& 0.00271447
\end{tabular}
\end{center} 
Now calculating the $Q_{0.05}$ and the probability value at $p_1 = 0.05$ and $p_2 = 0.5$ quantiles, we estimate the values of the other parameters. The value of the sigmoid parameter $\alpha$ is set at 10. The parameter values are shown in the following table. 
We are going to vary $(\psi, \psi^{0})$, $\gamma$, $\beta$ and r and see the effects on the strategy and portfolio wealth. The choices for these are also shown in the table below.

\begin{figure}[H]
\begin{subfigure}{.6\textwidth}
  \includegraphics[width=0.95\textwidth]{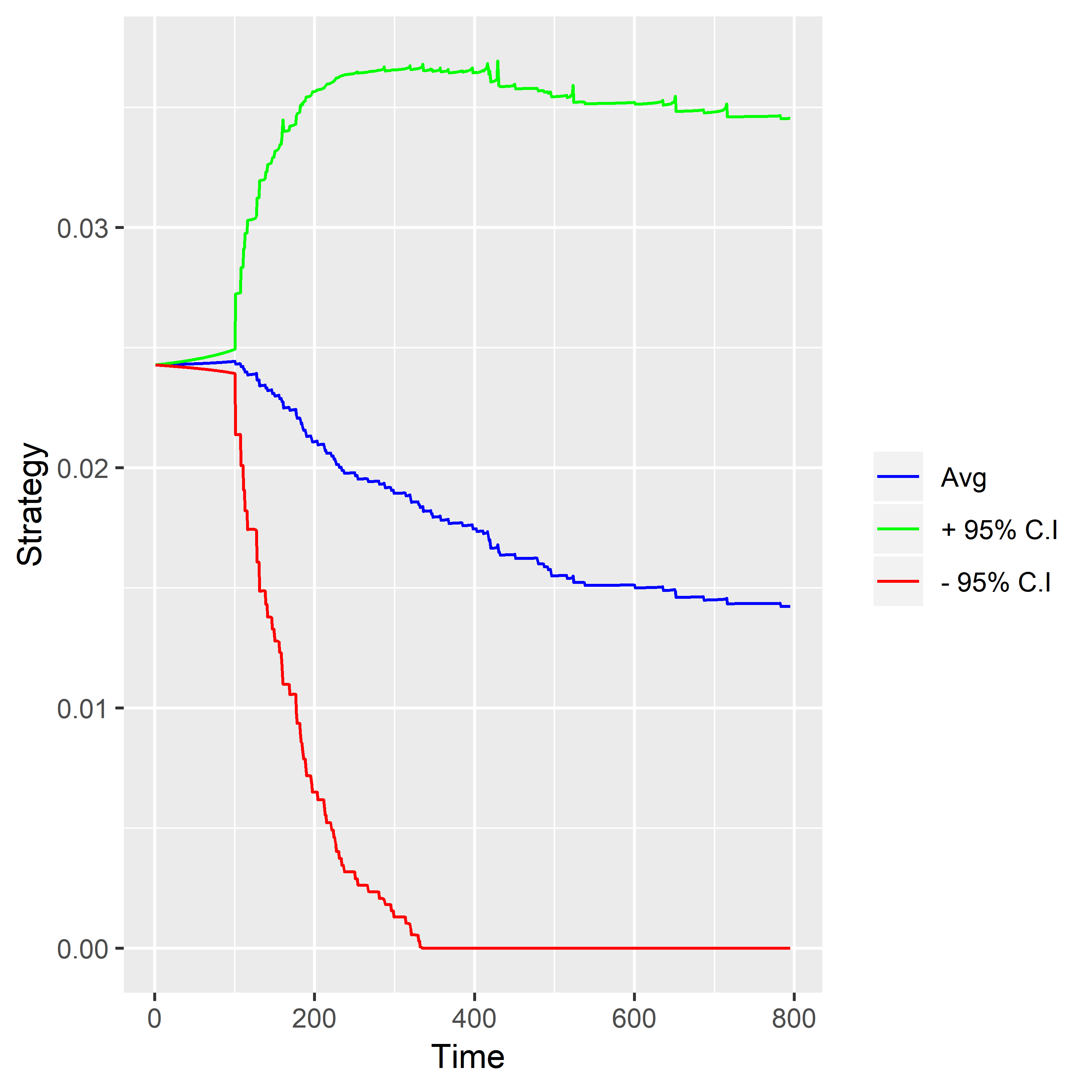}
\caption{r = 0.00014 }
\label{fig:20}
\end{subfigure}
\begin{subfigure}{.6\textwidth}
  \includegraphics[width=0.95\textwidth]{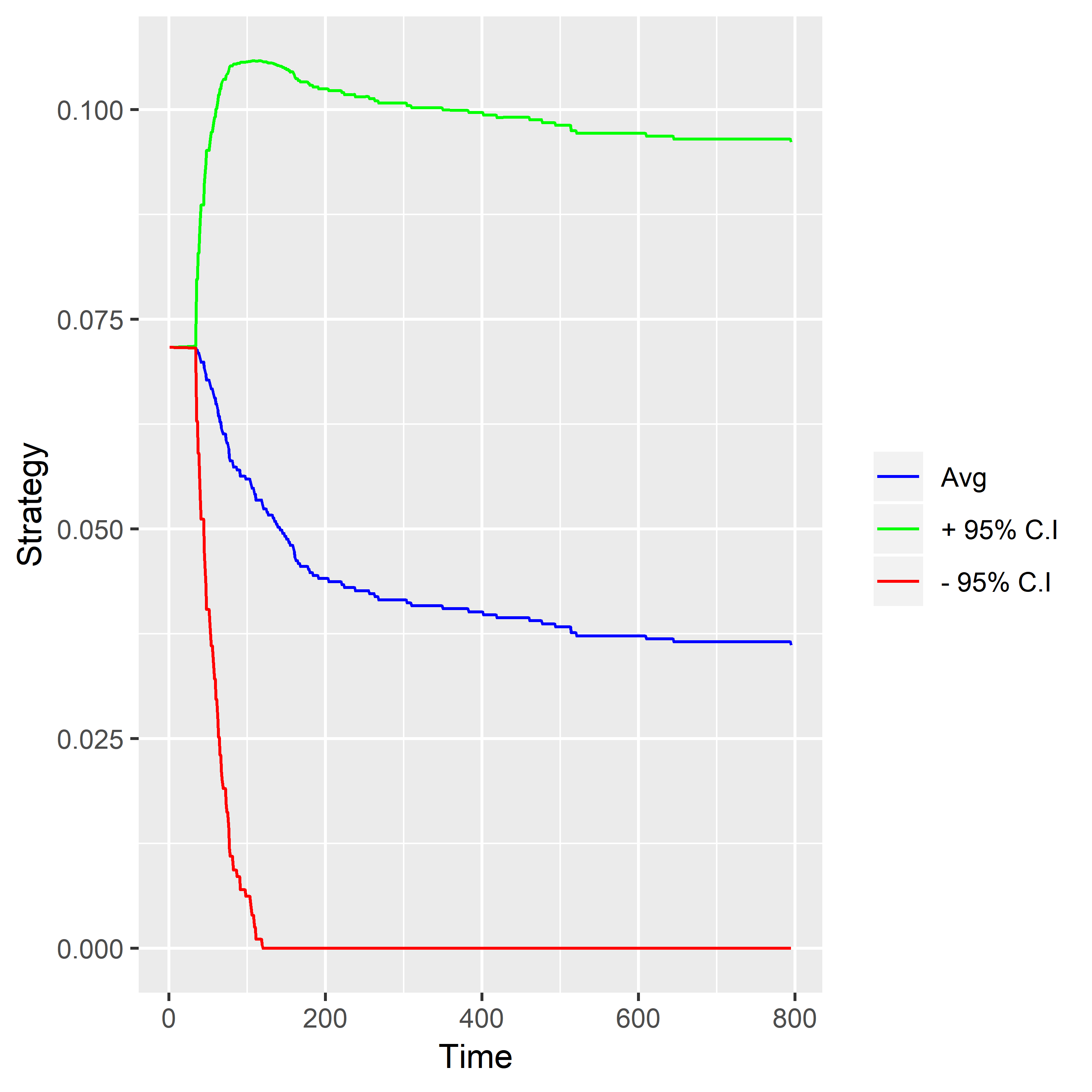}
   \caption{r = 0.0004 }
  \label{fig:16}
\end{subfigure}
\begin{subfigure}{.6\textwidth}
  \includegraphics[width=0.95\textwidth]{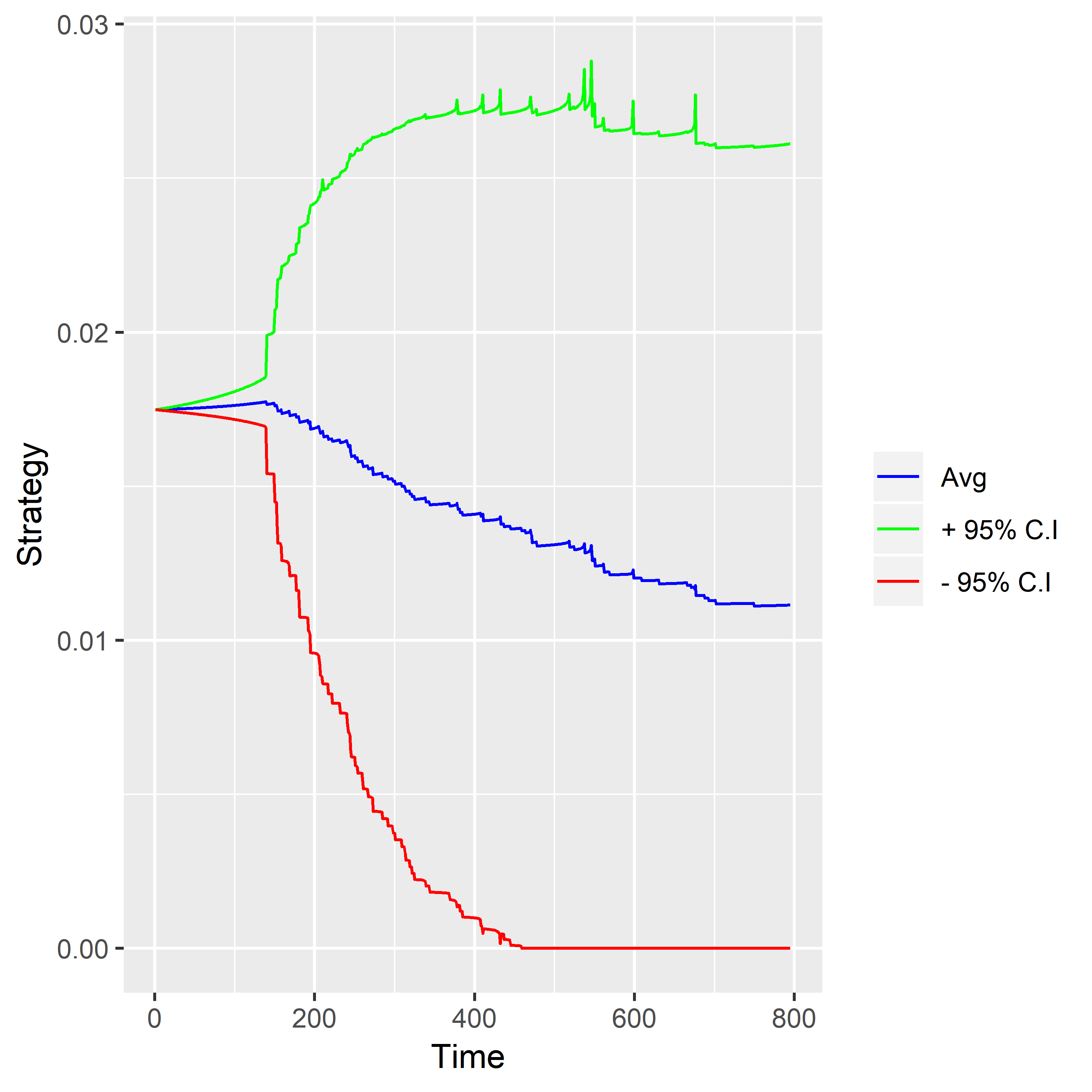}
\caption{r = 0.0001 }
  \label{fig:18}
\end{subfigure}
\caption{Average optimal strategy and its corresponding 95\% interval for different rates of interest (r).}
\label{fig:test}
\end{figure}

\begin{center}
\begin{tabular}{|ccl|}
     \hline $Q_{0.05}$ &=& 0.00077 \\ \hline $f_{0.05}$ &=& 47.63579 \\  \hline $f_{0.5}$ &=& 68.43975 \\  \hline $\alpha$ &=& 10 \\  \hline  $b_2$ &=& 0.00599\\ \hline $\rho$ &=& 0.22941 \\ \hline $\epsilon$ &=& 0.00001 \\  \hline $(\psi, \psi^{0})$ &=& (0.6, 0.8), (0.8, 0.6), (0.95, 0.312), (0.312, 0.95) \\ \hline 
     $\gamma$ &=& 0.1, 0.3, 0.5 \\ \hline $\beta$ &=& 0.001, 0.0005, 0.002 \\ \hline
     r &=& 0.0001, 0.00014, 0.0004 \\ \hline
\end{tabular}
\end{center}
Taking time period $t = 795$, $dt = 1$, $L_1 = 1$ and solving the above problem piece wise in time for random independent $dW^{(1)}$ and $dW^{(2)}$, we can make the approximation accurate while linearizing. Also we use the simulation from the Bi-variate Normal Distribution in R using Gibbs sampler to get the two Brownian motion. To illustrate our calculation steps, we take the following values $\psi = 0.6$, $\psi^{0} = 0.8$, $\gamma = 0.3$ and $\beta = 0.001$. Putting these values in \eqref{final} for r = 0.00014 we get the figure (\ref{fig:20}) for the optimal strategy.

\noindent  We generate 200 sample paths each for all the configurations and then calculate the average (point-wise, call it $\mu(t)$) for all the sample paths. Once the average is calculated, we calculate the standard deviation (similarly, $\sigma(t)$) of the paths. Considering normal distribution and using the z-score value which corresponds to $(\mu(t) \pm 1.96\times \sigma(t))$ we find 95\% confidence interval for optimal strategy from all these random samples of the Brownian motion. 

\begin{figure}[H]
\begin{subfigure}{.6\textwidth}
  \includegraphics[width=0.95\textwidth]{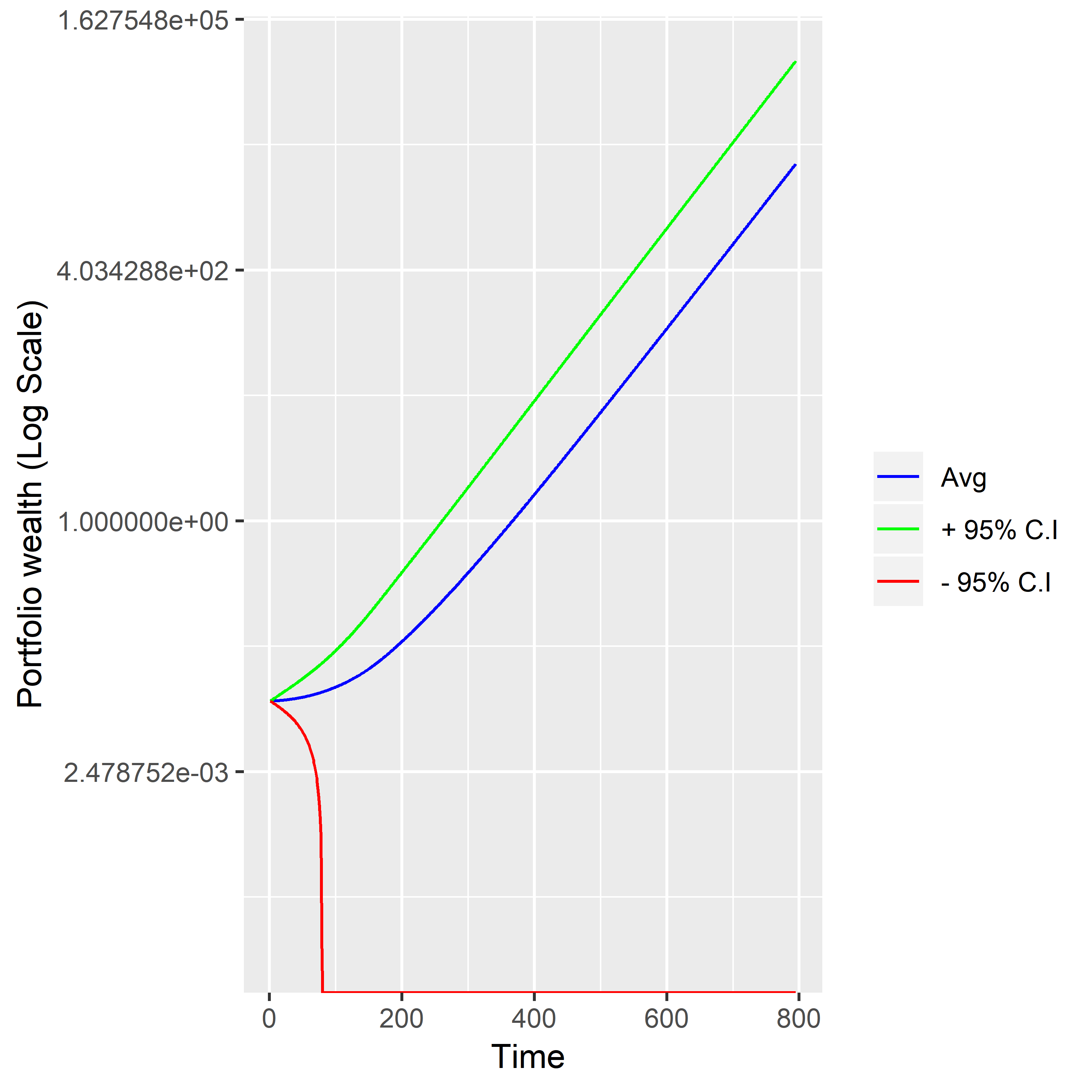}
\caption{r = 0.00014 }
\label{fig:21}
\end{subfigure}
\begin{subfigure}{.6\textwidth}
  \includegraphics[width=0.95\textwidth]{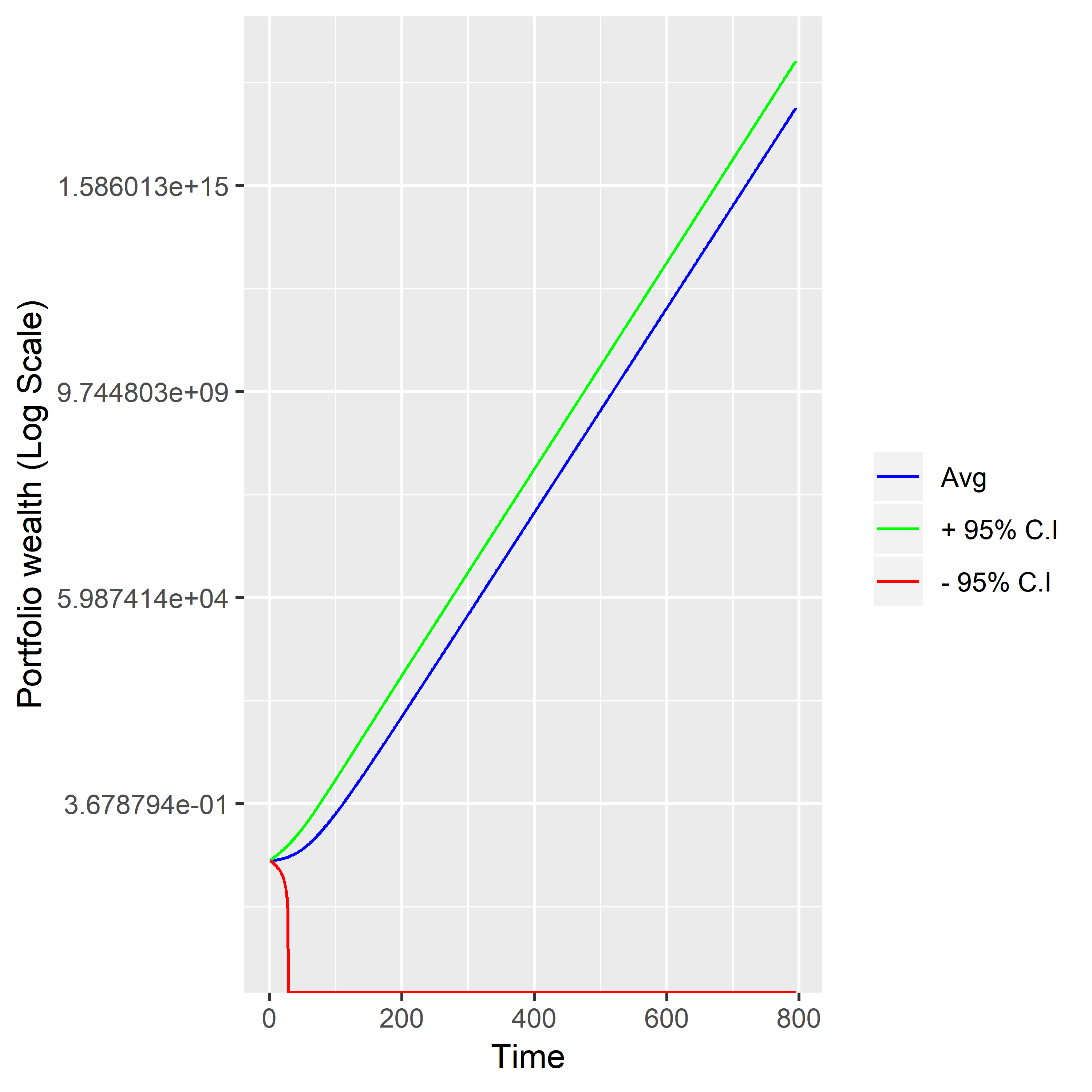}
   \caption{r = 0.0004 }
  \label{fig:17}
\end{subfigure}
\begin{subfigure}{.6\textwidth}
  \includegraphics[width=0.95\textwidth]{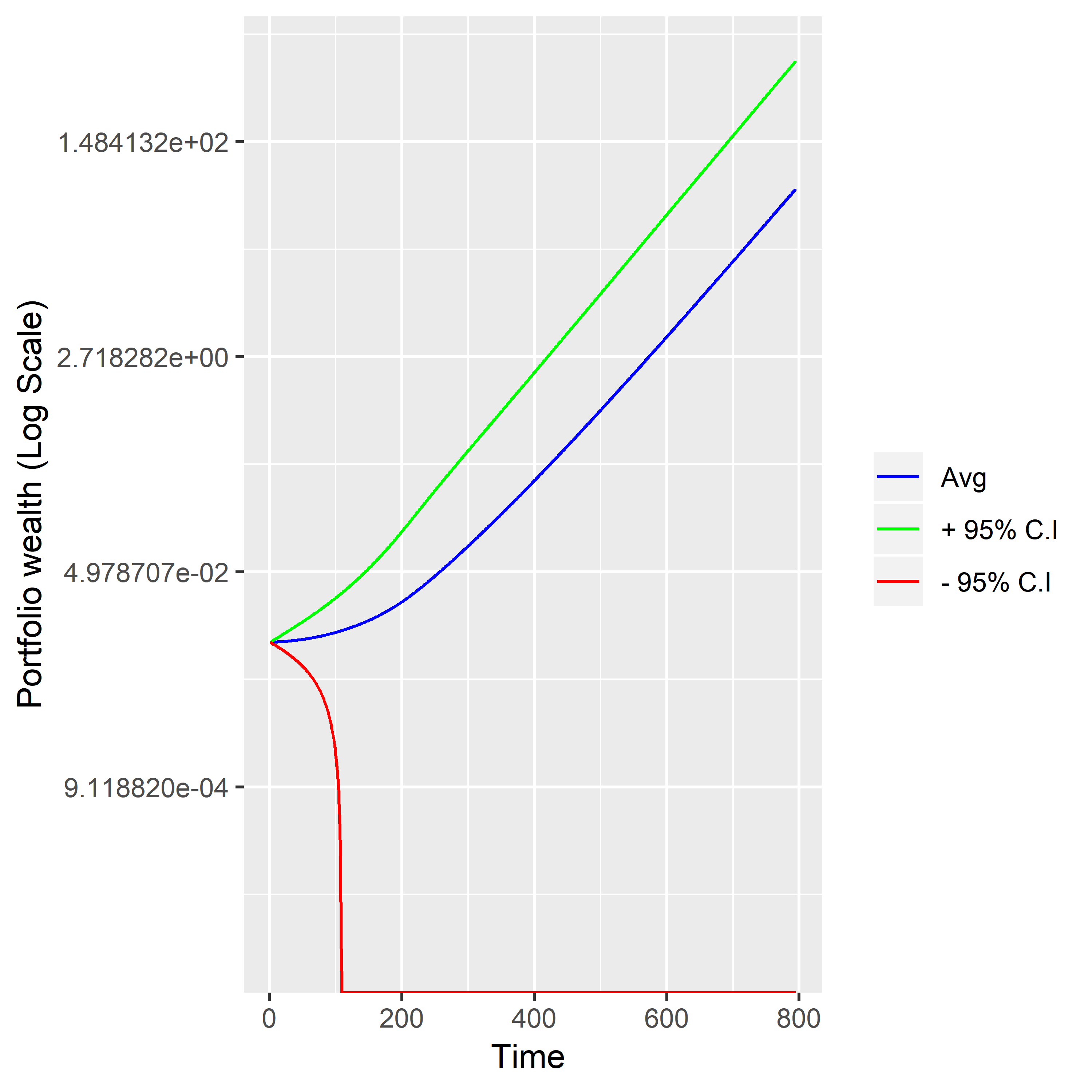}
\caption{r = 0.0001 }
  \label{fig:19}
\end{subfigure}
\caption{Portfolio wealth and its corresponding 95\% interval for different rates of interest (r).}
\label{fig:test}
\end{figure}

\begin{itemize}
    \item 
Firstly, keeping other things fixed, we vary the rate of interest (r = 0.0001, 0.00014 and 0.0004). When r = 0.0004 as used in \eqref{approx} and  drawing the plot we see that it lies between the range (0.074, 0.038) as shown in the figure (\ref{fig:16}).  Similarly we can show that the portfolio wealth goes up to an order of $10^{15}$ as shown in the figure (\ref{fig:17}). The rate of change of the optimal strategy and the portfolio wealth is maximum as compared to the other two interest rates.

The 95\% confidence interval for optimal strategy for all the random samples of the Brownian motion for r = 0.0001 are used in \eqref{approx} shows that the strategy lies in the range (0.0175, 0.012) and the gradient of decrease is minimum as compared to the other two rates of interest as shown in the figure (\ref{fig:18}).  Similarly we can show that the portfolio wealth goes up to the range of $10^{2}$ as shown in the figure (\ref{fig:19}).

Finally, the 95\% confidence interval for optimal strategy for all the random samples of the Brownian motion for r = 0.00014 are used in \eqref{approx} as mentioned previously. As time increases the optimal strategy gradually decreases and it lies between (0.024, 0.015) refer Figure (\ref{fig:20}).The portfolio wealth increases up to a range of $10^{5}$ during this time period as shown in the figure (\ref{fig:21}). From the above discussion and plots we can conclude that as the rate of interest increases the amount of wealth to be invested on stock decreases. This is intuitive as when the risk less asset becomes more attractive, it makes sense to invest more in it so that risk is minimised without compromising too much on return.
\begin{center}
\textbf{Figure 1 and Figure 2 should be placed here}
\end{center}
\item
Secondly, we are going to consider the scenario where weights $(\psi, \psi^{0})$ are modified keeping other parameters constant. As we change the weights of the optimizing function from 0.8 to 0.6 and the constraint weights from 0.6 to 0.8 and plot the optimal strategy for r = 0.00014, refer figure (\ref{fig:20_1}) and figure (\ref{fig:14}). When we are using the weights for objective function as 0.312 and the constraint as 0.95, refer figure (\ref{fig:22}) and when we interchange the weights accordingly refer figure (\ref{fig:24}). From these plots we can see that as we increase the objective weights and give less weightage to subject constraint then the amount to be invested in stocks should be more. Again, this result is in line with financial intuition.

Looking in to the portfolio wealth as we change the weights of the optimizing function from 0.6 to 0.8 and the constraint weights from 0.8 to 0.6 and plot the wealth for r = 0.00014, refer figure (\ref{fig:21_1}) and figure (\ref{fig:15}). When we are using the weights for objective function as 0.312 and the constraint as 0.95, refer figure (\ref{fig:23}) and when we interchange the weights accordingly, refer figure (\ref{fig:25}). From these graphs we can see that as we increase the objective weights and give less weightage to the constraint, then portfolio wealth increases. Again, giving less weightage to the constraint means taking more risk (compromising on the VaR) in order to increase the expected (median) gain. So, naturally, in such a case the portfolio wealth increases.

\begin{figure}[H]
\centering
\begin{subfigure}{.6\textwidth}
  \centering
  \includegraphics[width=0.95\textwidth]{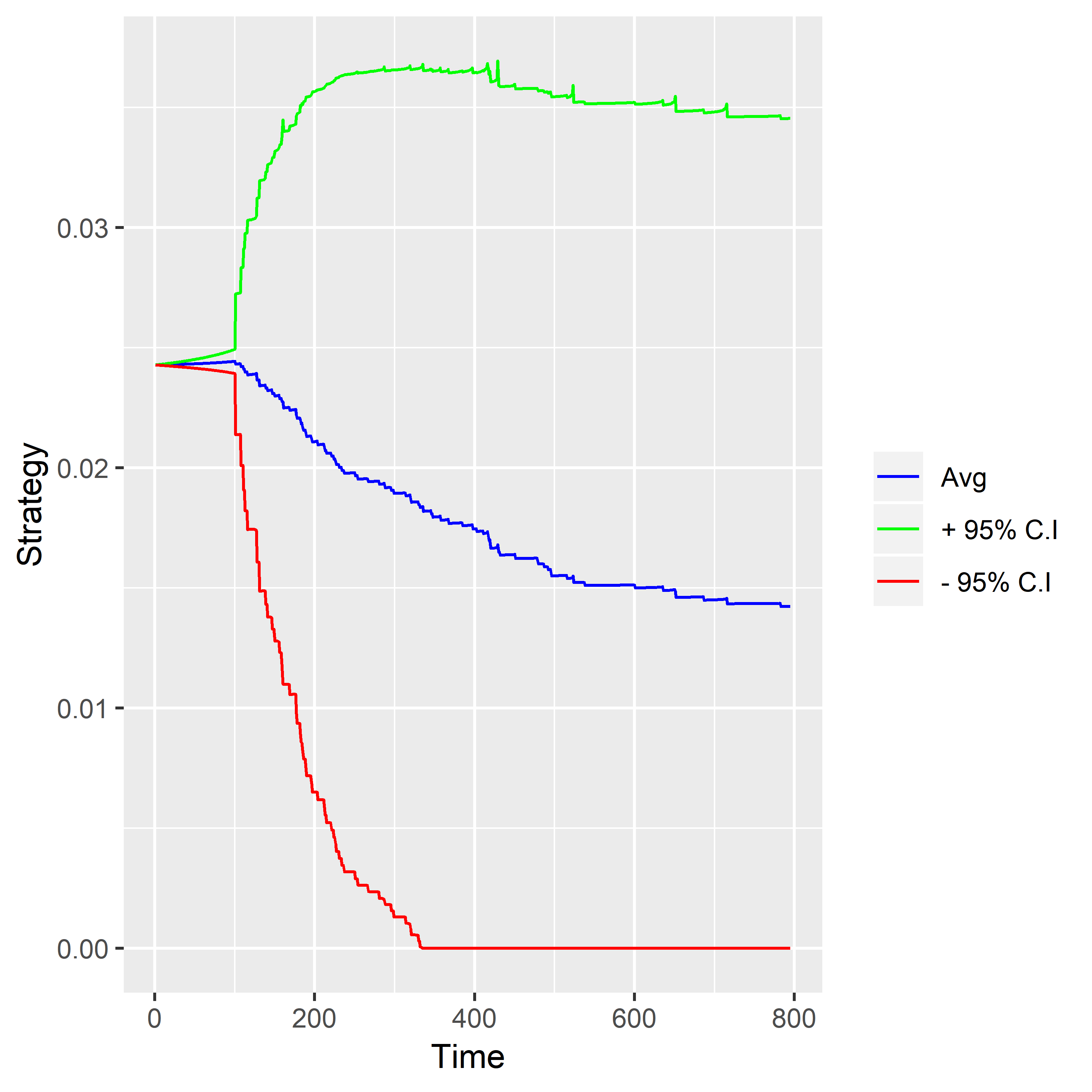}
   \caption{$\psi$ = 0.6, $\psi^{0}$ = 0.8 }
  \label{fig:20_1}
\end{subfigure}%
\begin{subfigure}{.6\textwidth}
  \centering
  \includegraphics[width=0.95\textwidth]{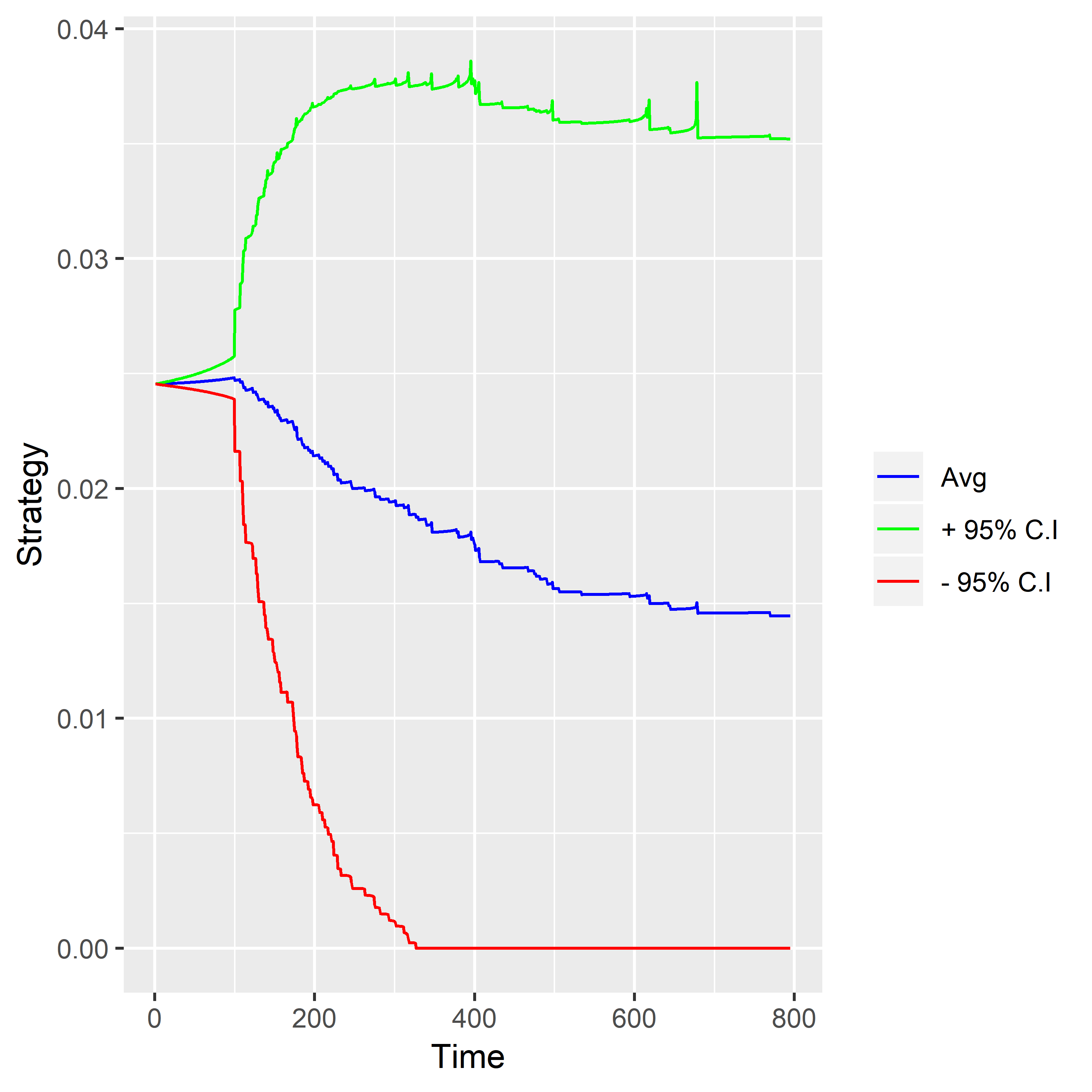}
\caption{$\psi$ = 0.8, $\psi^{0}$ = 0.6 }
  \label{fig:14}
\end{subfigure}
\begin{subfigure}{.6\textwidth}
  \centering
  \includegraphics[width=0.95\textwidth]{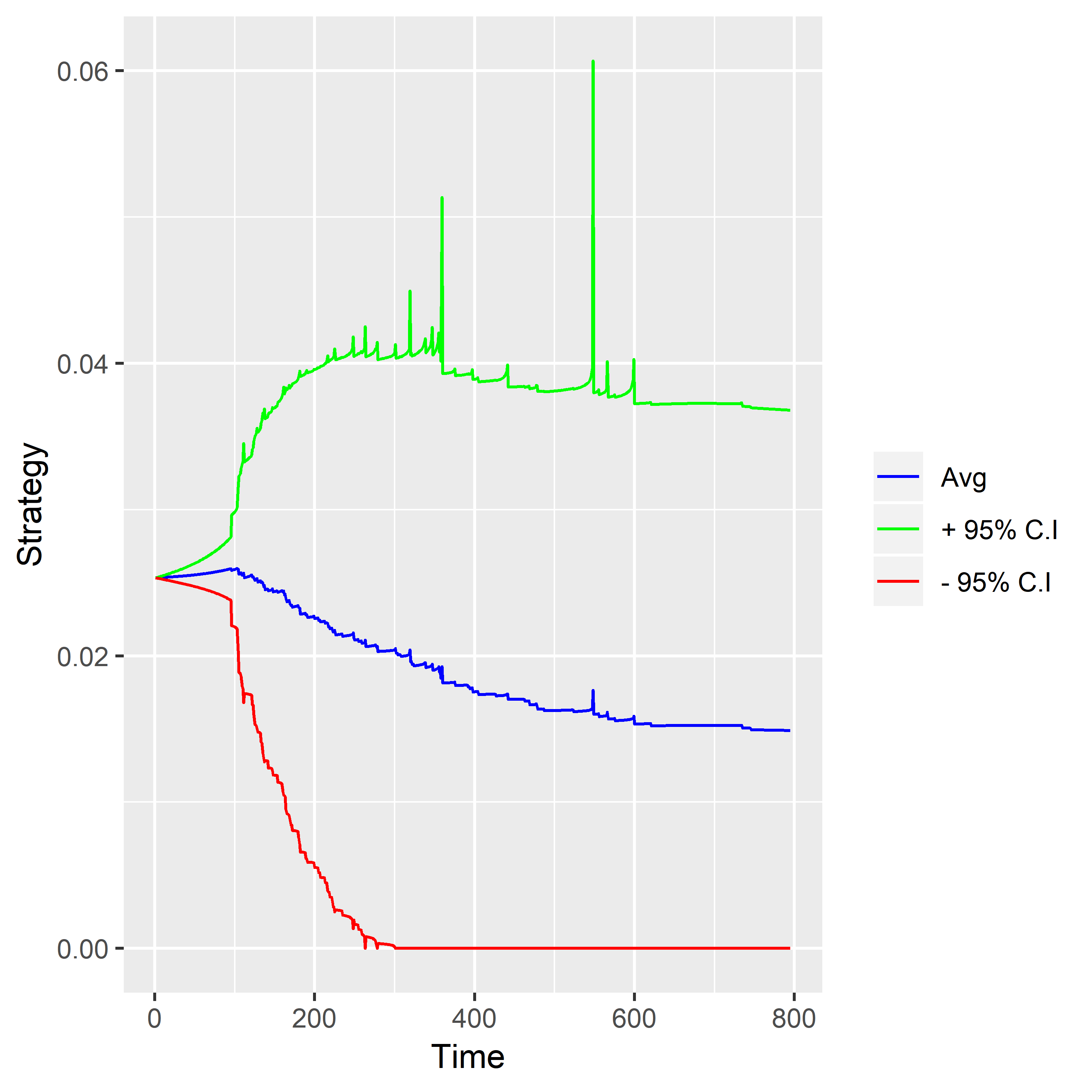}
   \caption{$\psi$ = 0.95, $\psi^{0}$ = 0.312 }
  \label{fig:22}
\end{subfigure}%
\begin{subfigure}{.6\textwidth}
  \centering
  \includegraphics[width=0.95\textwidth]{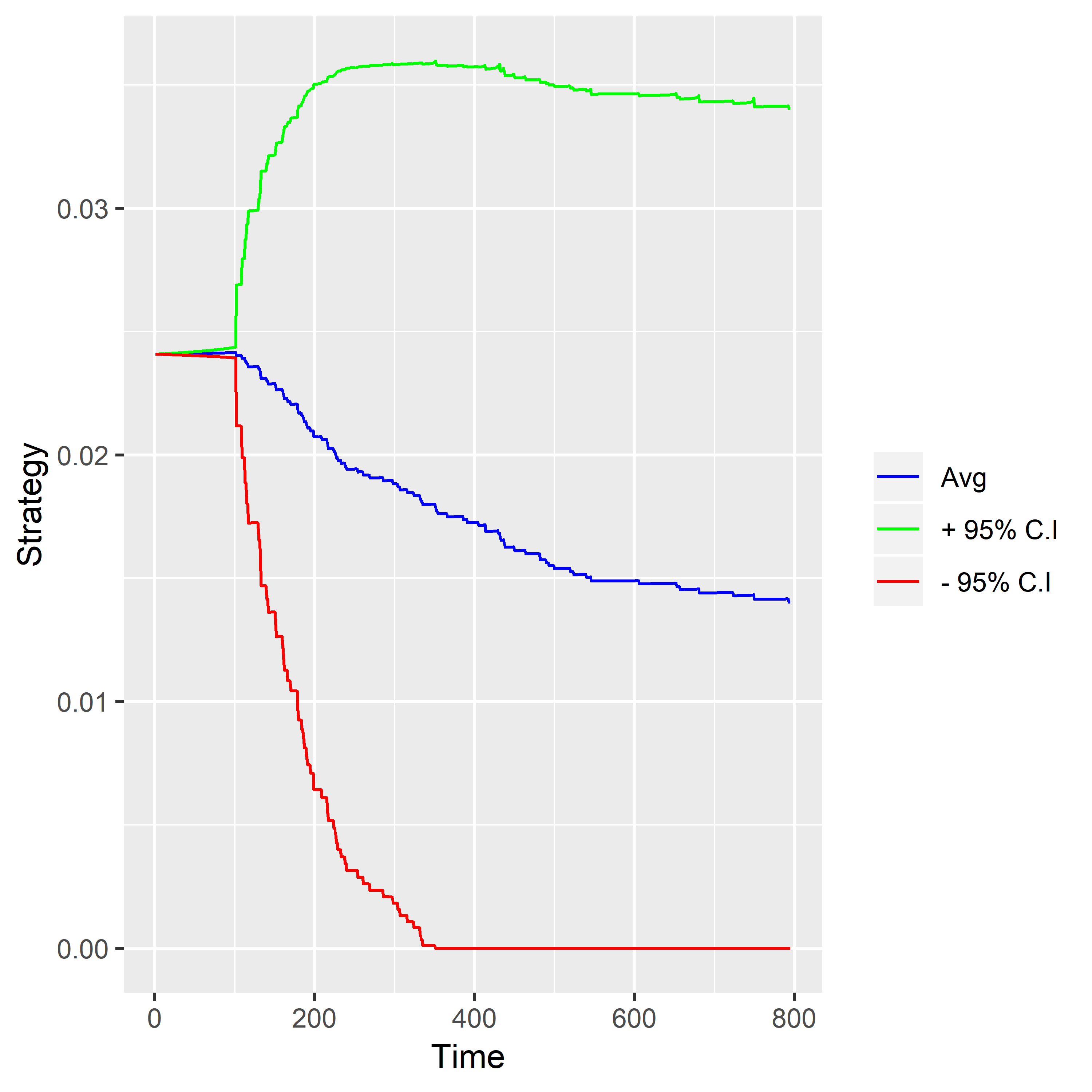}
\caption{$\psi$ = 0.312, $\psi^{0}$ = 0.95 }
  \label{fig:24}
\end{subfigure}
\caption{Average optimal strategy and its corresponding 95\% interval r = 0.00014, $\gamma$ = 0.3, $\beta$ = 0.001}
\label{fig:test}
\end{figure}

\begin{figure}[H]
\centering
\begin{subfigure}{.6\textwidth}
  \centering
  \includegraphics[width=0.95\textwidth]{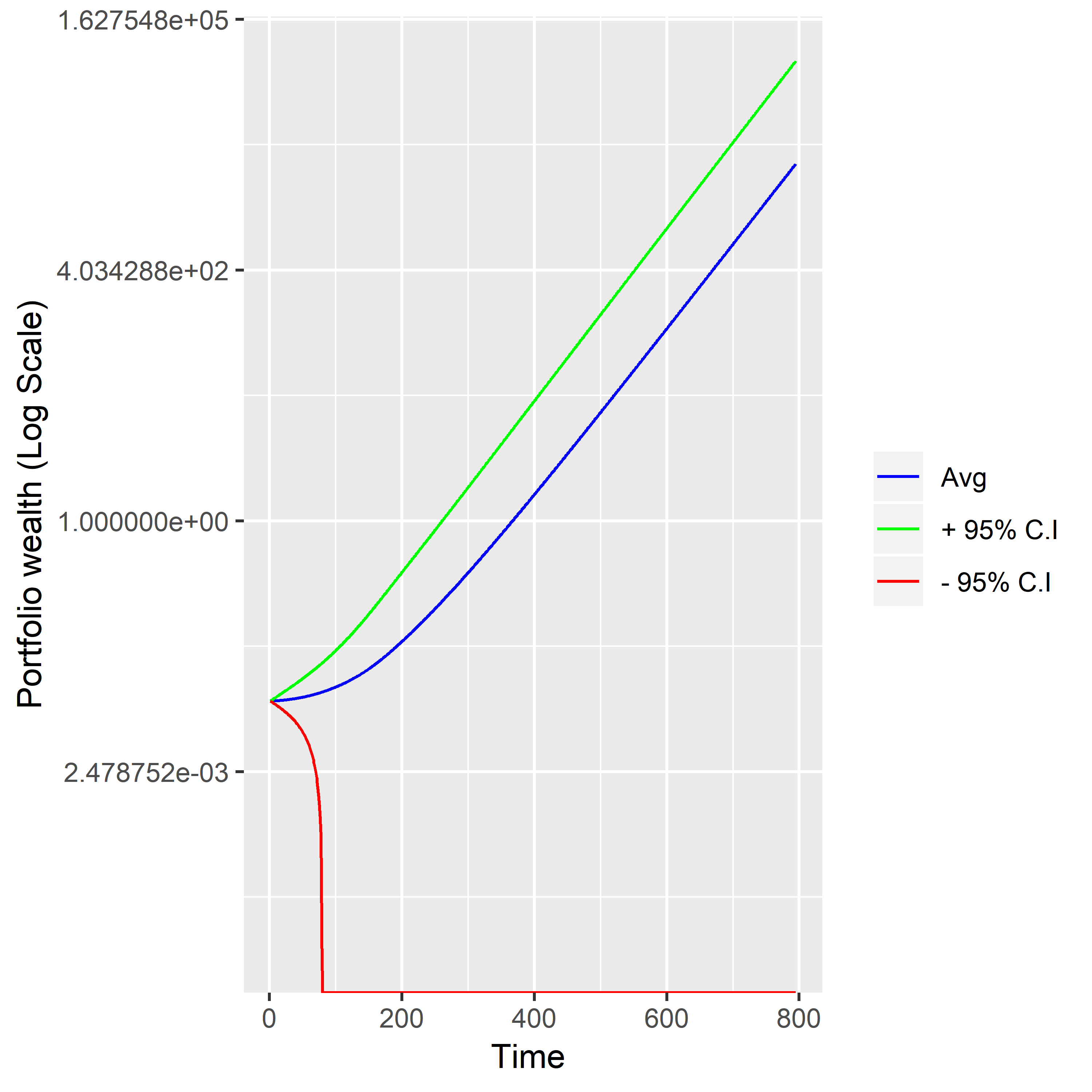}
   \caption{$\psi$ = 0.6, $\psi^{0}$ = 0.8 }
  \label{fig:21_1}
\end{subfigure}%
\begin{subfigure}{.6\textwidth}
  \centering
  \includegraphics[width=0.95\textwidth]{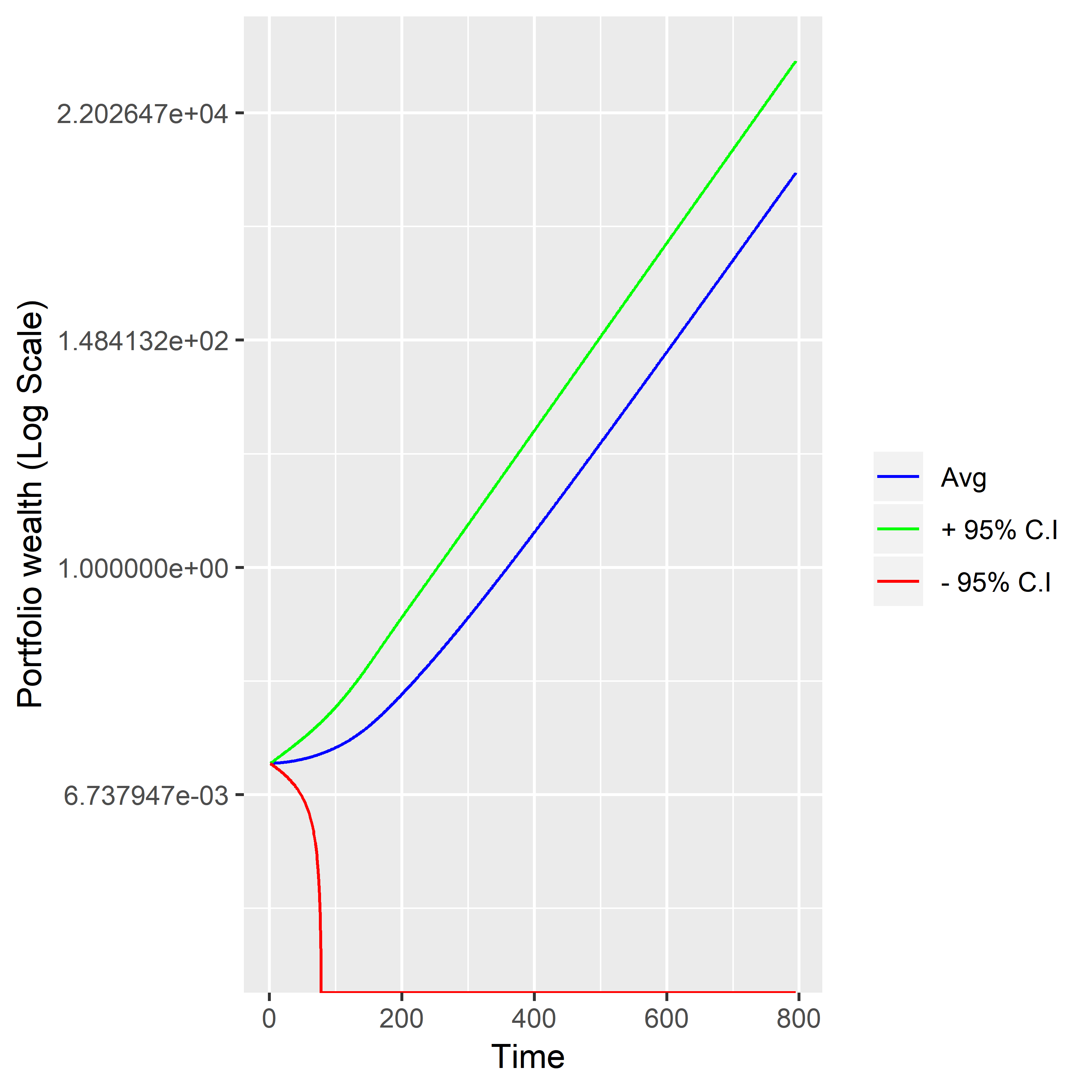}
\caption{$\psi$ = 0.8, $\psi^{0}$ = 0.6 }
  \label{fig:15}
\end{subfigure}
\begin{subfigure}{.6\textwidth}
  \centering
  \includegraphics[width=0.95\textwidth]{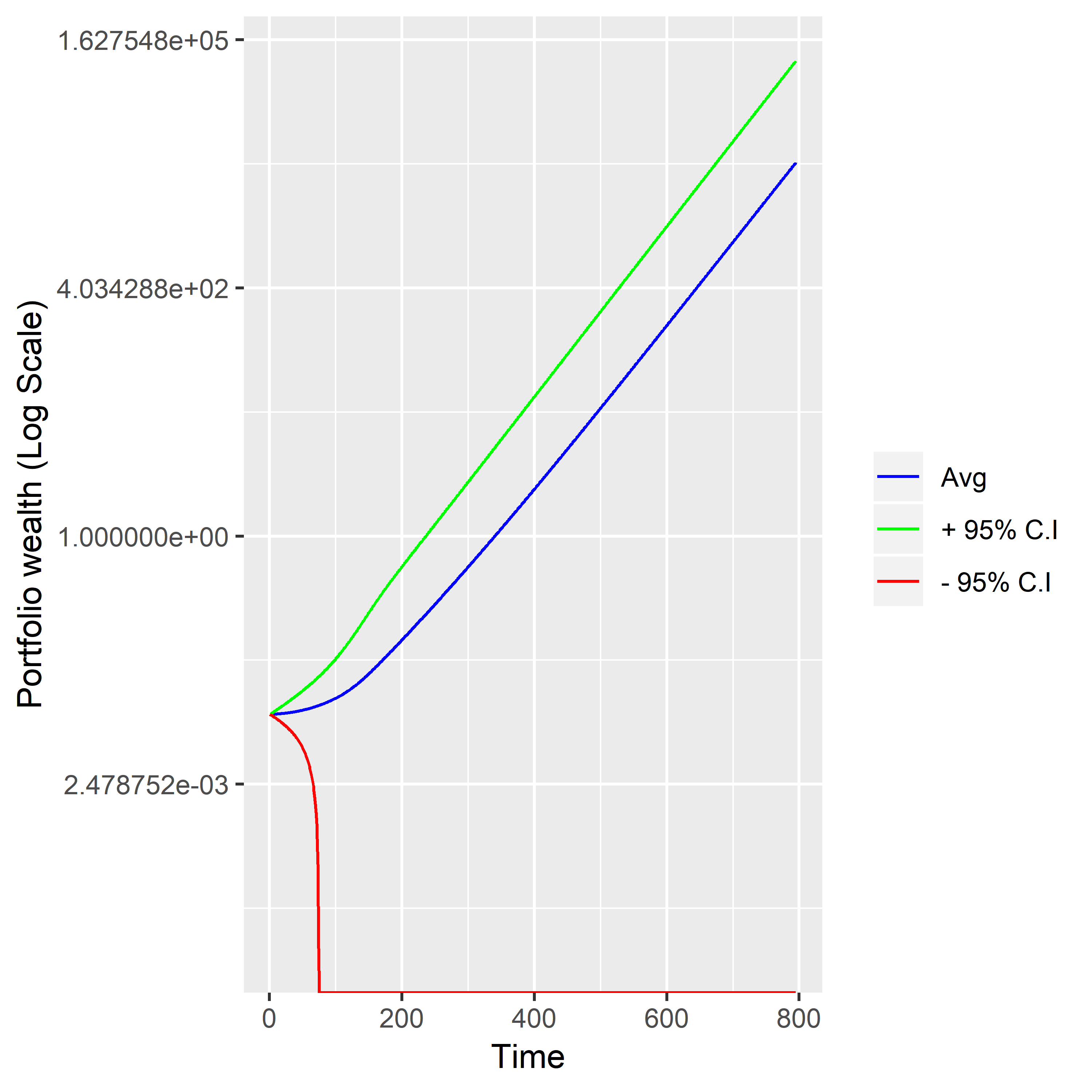}
   \caption{$\psi$ = 0.95, $\psi^{0}$ = 0.312 }
  \label{fig:23}
\end{subfigure}%
\begin{subfigure}{.6\textwidth}
  \centering
  \includegraphics[width=0.95\textwidth]{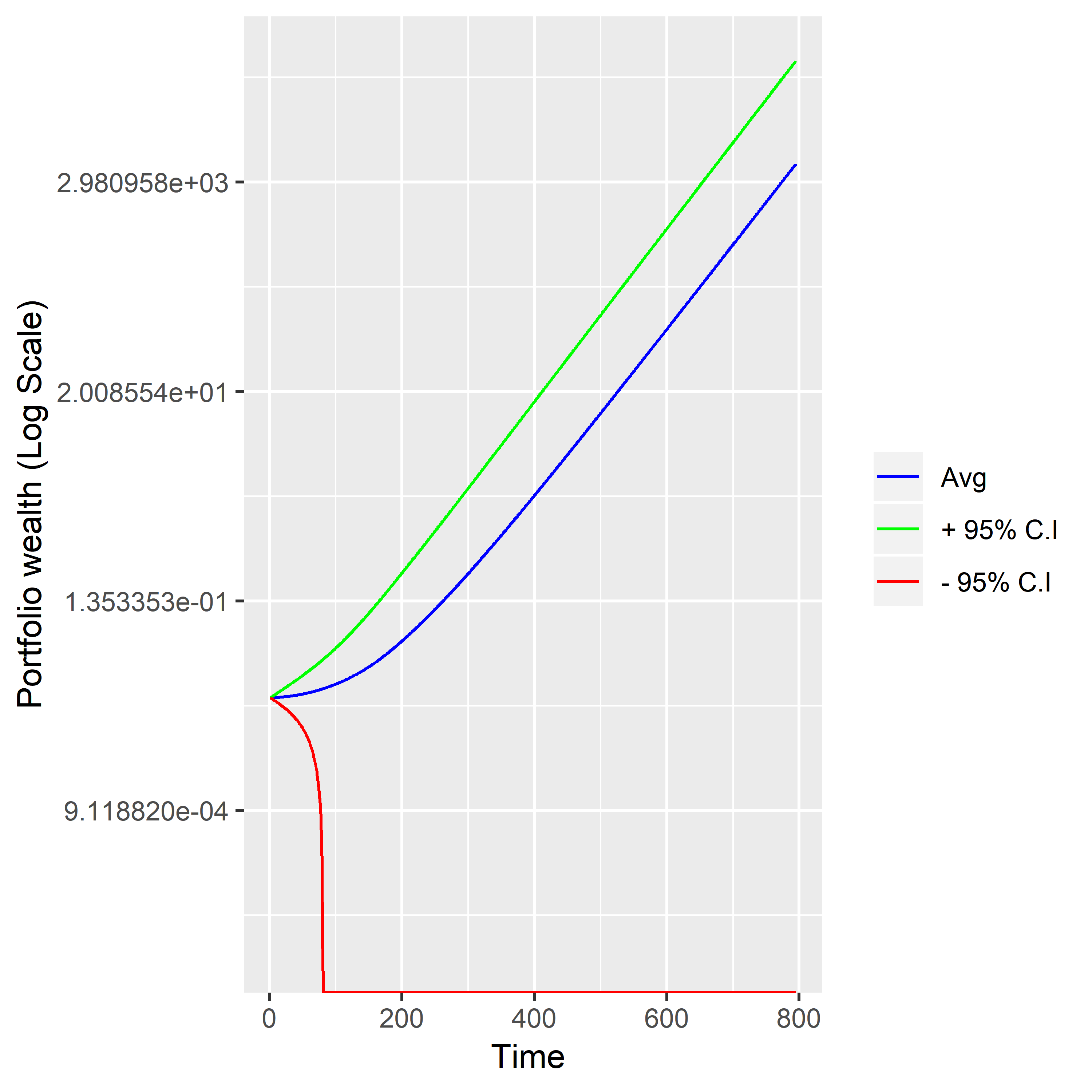}
\caption{$\psi$ = 0.312, $\psi^{0}$ = 0.95 }
  \label{fig:25}
\end{subfigure}
\caption{Average portfolio wealth and its corresponding 95\% interval r = 0.00014, $\gamma$ = 0.3, $\beta$ = 0.001}
\label{fig:test}
\end{figure}
\begin{center}
\textbf{Figure 3 and Figure 4 should be placed here}
\end{center}
\item
Next, we would like to see the effect of the risk aversion parameter $\gamma$ keeping all other factors constant. We plot the optimal strategy for $\gamma = 0.3$ [refer figure (\ref{fig:20_2})], then we decrease $\gamma$ from 0.3 to 0.1 and plot the optimal strategy [refer figure (\ref{fig:26})]. Similarly, we increase $\gamma$ from 0.3 to 0.5 and plot the optimal strategy [refer figure (\ref{fig:28})]. We conclude that all the three graphs are similar so there is no specific change in the optimal strategy if we change the risk aversion parameter $\gamma$.

\begin{figure}[H]
\begin{subfigure}{.6\textwidth}
  \includegraphics[width=0.95\textwidth]{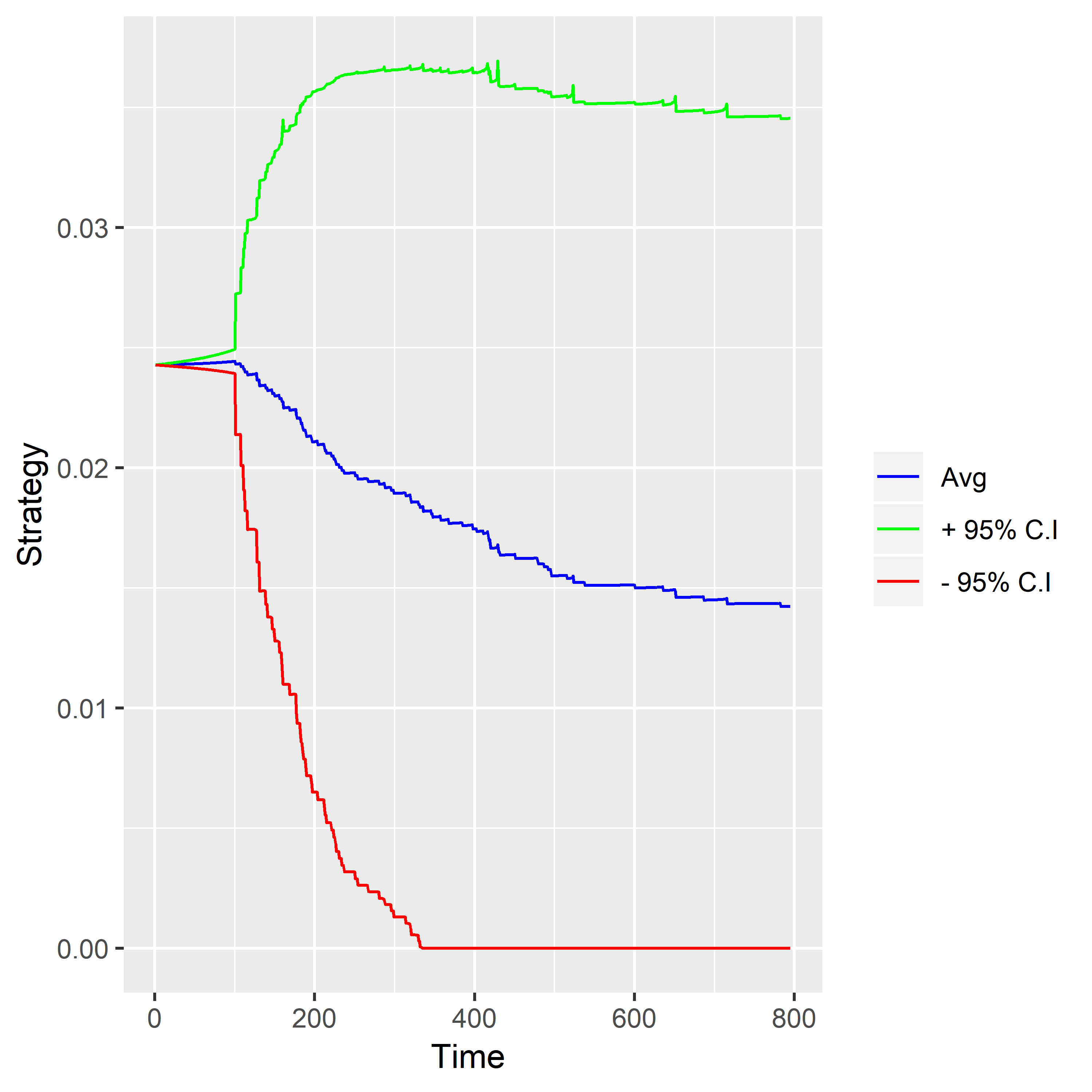}
   \caption{$\gamma = 0.3$ }
  \label{fig:20_2}
\end{subfigure}
\begin{subfigure}{.6\textwidth}
  \includegraphics[width=0.95\textwidth]{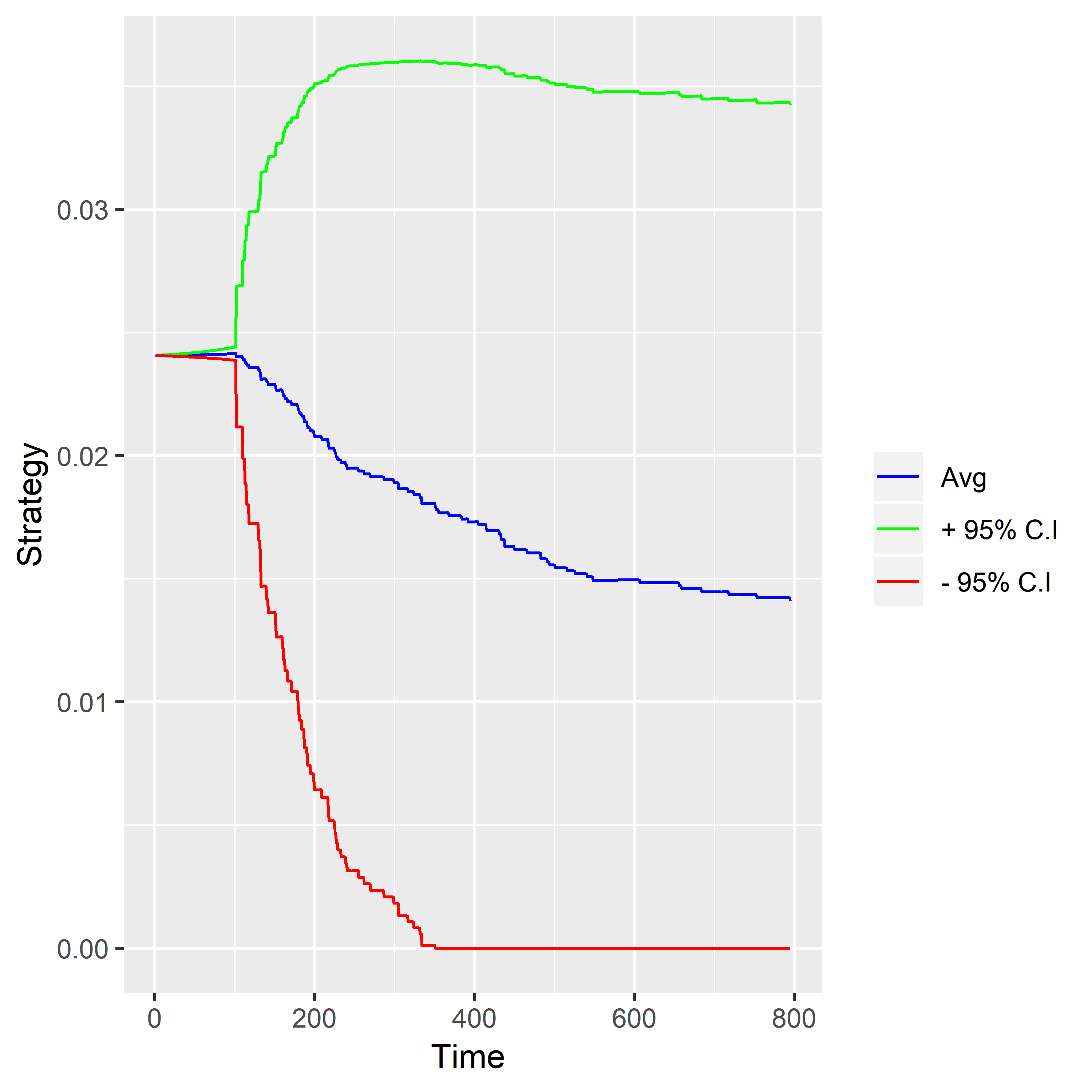}
\caption{$\gamma = 0.1$ }
  \label{fig:26}
\end{subfigure}
\begin{subfigure}{.6\textwidth}
  \includegraphics[width=0.95\textwidth]{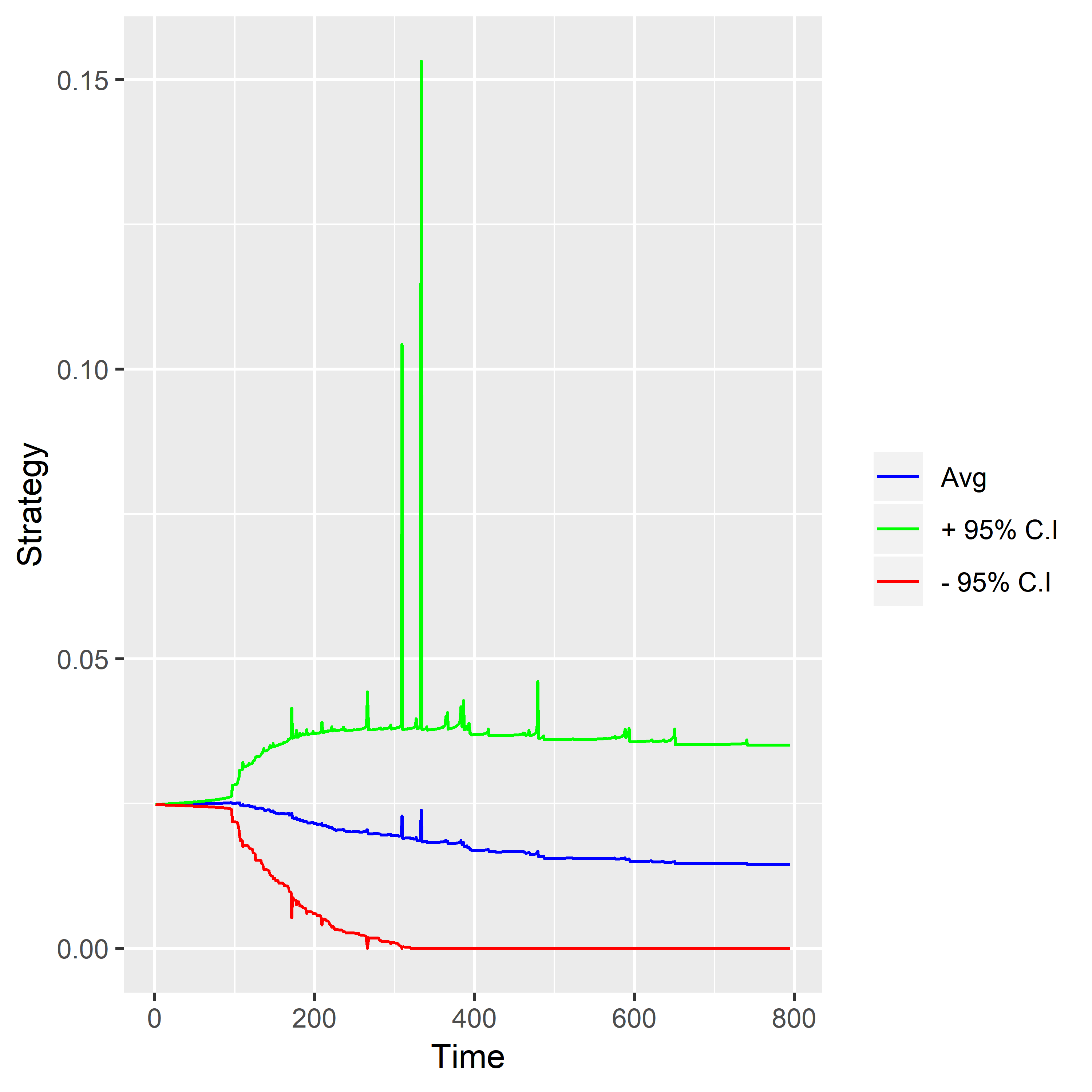}
   \caption{$\gamma = 0.5$ }
  \label{fig:28}
\end{subfigure}
\caption{Average optimal strategy and its corresponding 95\% interval r = 0.00014, $\psi$ = 0.6, $\psi^{0}$ = 0.8, $\beta$ = 0.001}
\label{fig:test}
\end{figure}

If we plot the wealth for different levels of risk aversion, as illustrated by the three values of $\gamma$ we have chosen, we can see that for all the cases the wealth accumulates in the same way, [refer figure (\ref{fig:21_2}, \ref{fig:27} and \ref{fig:29})].
\begin{figure}[H]
\begin{subfigure}{.6\textwidth}
  \centering
  \includegraphics[width=0.95\textwidth]{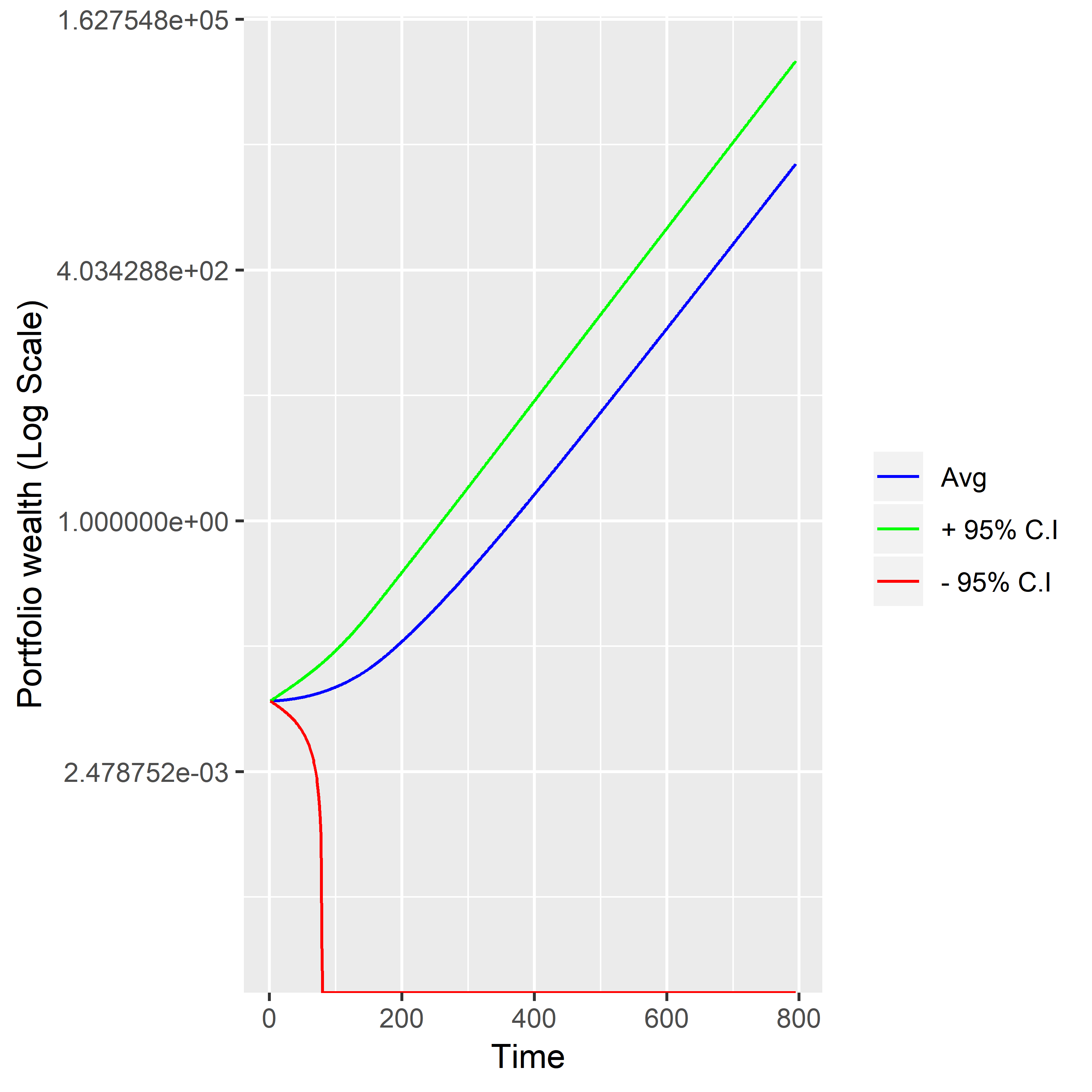}
   \caption{$\gamma = 0.3$ }
  \label{fig:21_2}
\end{subfigure}
\begin{subfigure}{.6\textwidth}
  \includegraphics[width=0.95\textwidth]{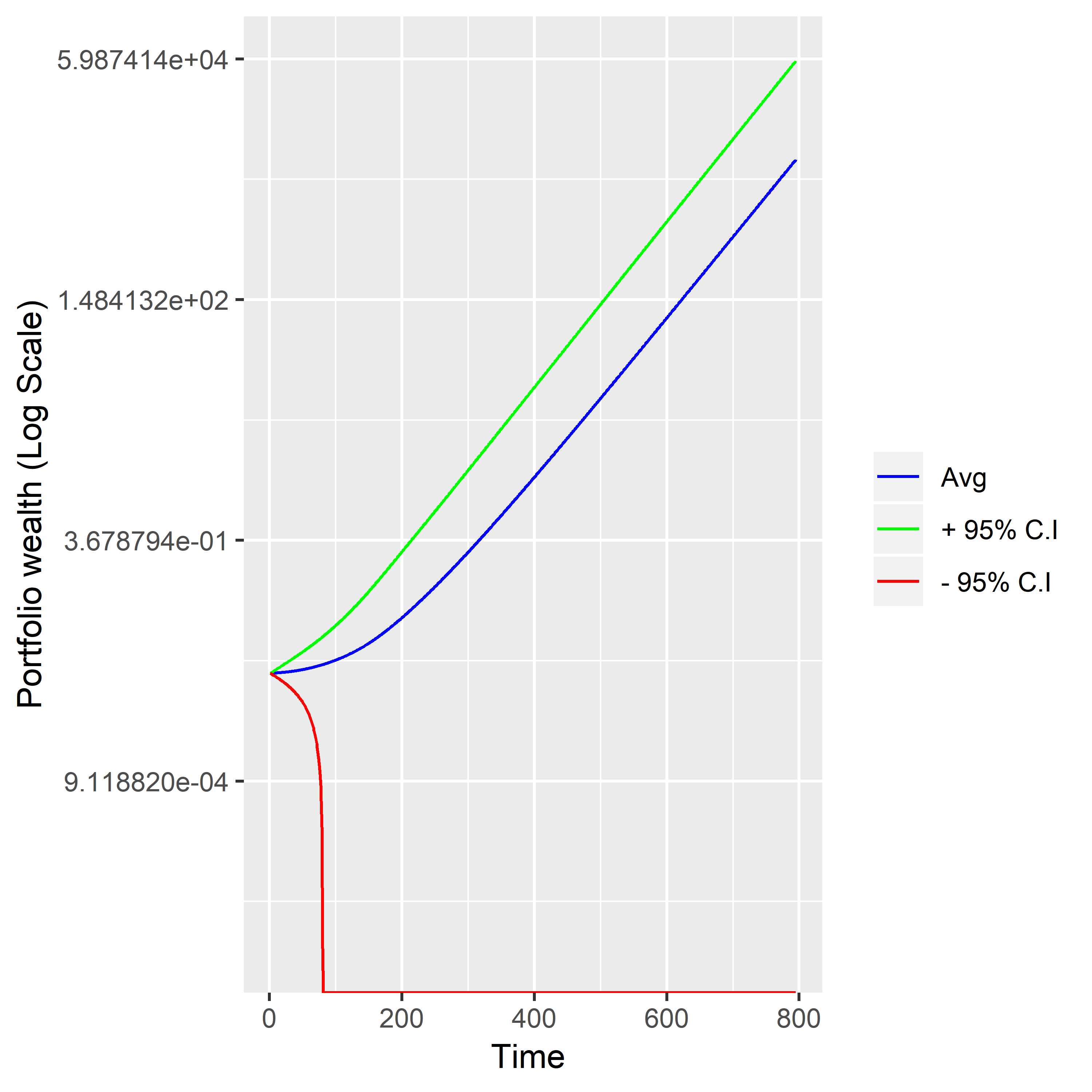}
\caption{$\gamma = 0.1$ }
  \label{fig:27}
\end{subfigure}
\begin{subfigure}{.6\textwidth}
  \includegraphics[width=0.95\textwidth]{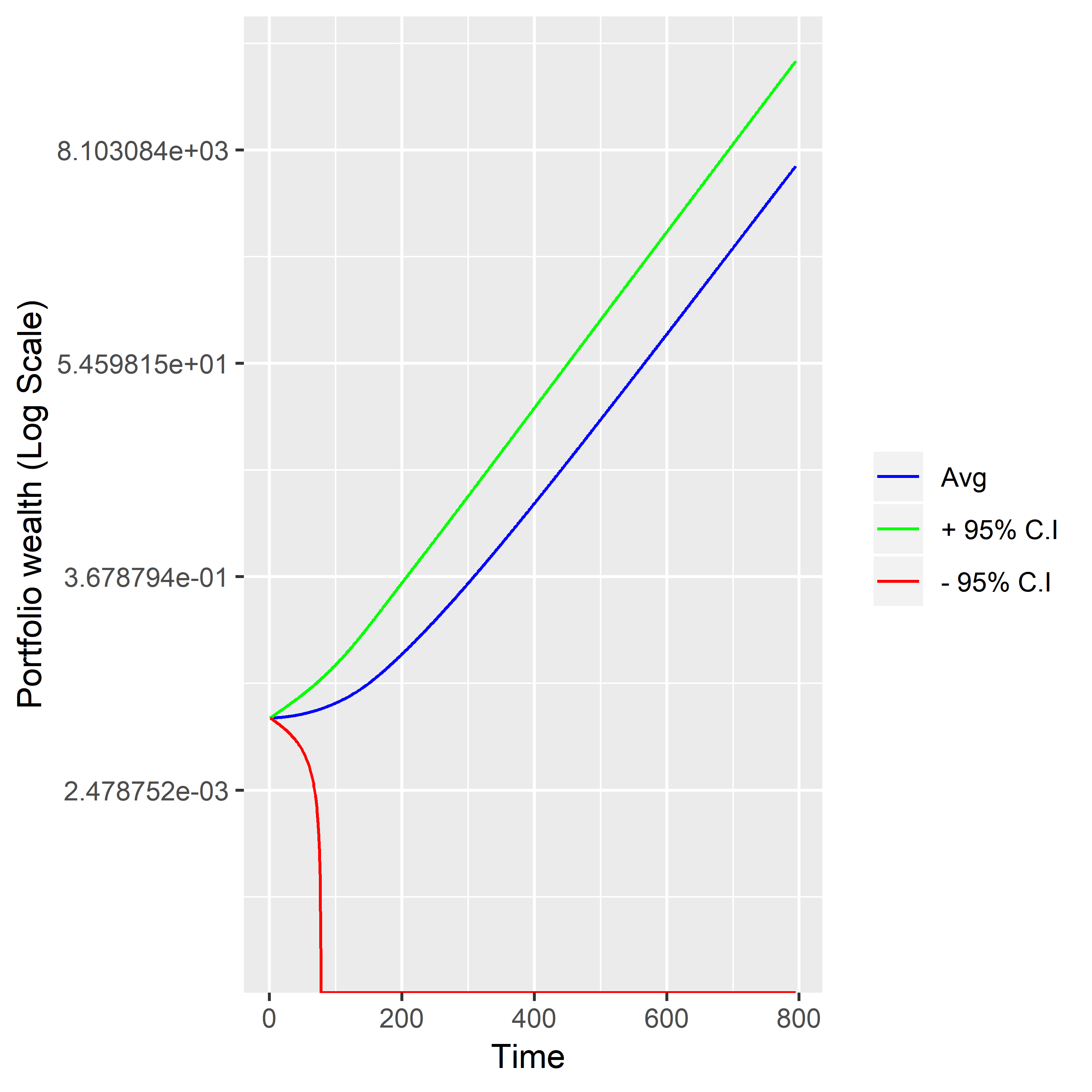}
   \caption{$\gamma = 0.5$ }
  \label{fig:29}
\end{subfigure}
\caption{Average wealth and its corresponding 95\% interval r = 0.00014, $\psi$ = 0.6, $\psi^{0}$ = 0.8, $\beta$ = 0.001}
\label{fig:test}
\end{figure}
\begin{center}
\textbf{Figure 5 and Figure 6 should be placed here}
\end{center}
\item
Finally, we see the effect of change in the time discounting, $\beta$, keeping all other factors constant. We plot optimal strategy for $\beta = 0.001$ [refer figure (\ref{fig:20_3})], then we decrease $\beta$ from 0.001 to 0.0005 and plot the optimal strategy [refer figure (\ref{fig:20_3}) and (\ref{fig:30})]. Similarly we increase $\beta$ from 0.001 to 0.002 [refer figure (\ref{fig:32})]. In each of the situation all the three graphs are similar so we can conclude that the change in the time discounting factor $\beta$ doesn't change the optimal strategy. But the variability of the optimal strategy, as illustrated by the width of the confidence interval, increases as we increase $\beta$.

\begin{figure}[H]
\begin{subfigure}{.6\textwidth}
  \centering
  \includegraphics[width=0.95\textwidth]{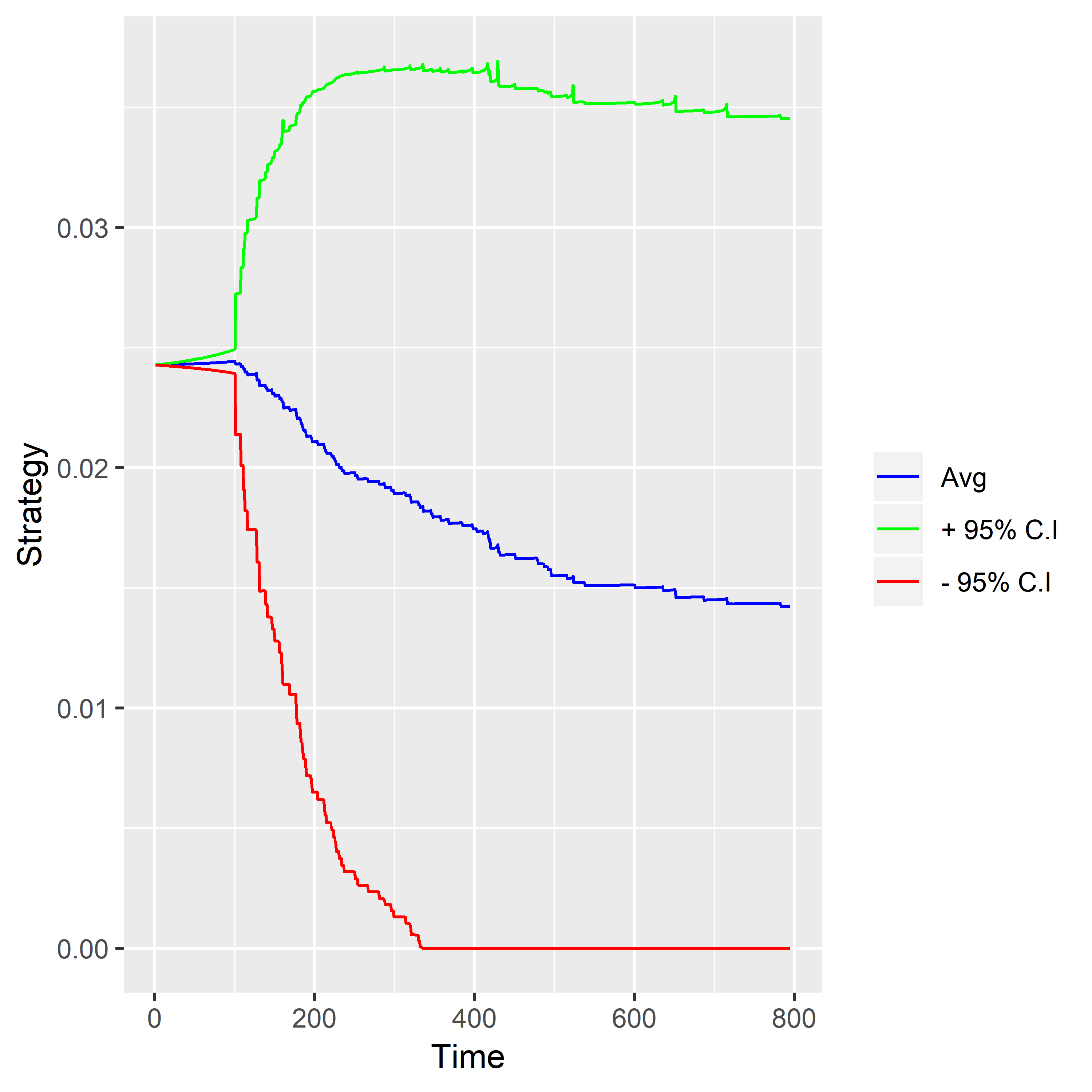}
   \caption{$\beta = 0.001$ }
  \label{fig:20_3}
\end{subfigure}
\begin{subfigure}{.6\textwidth}
  \includegraphics[width=0.95\textwidth]{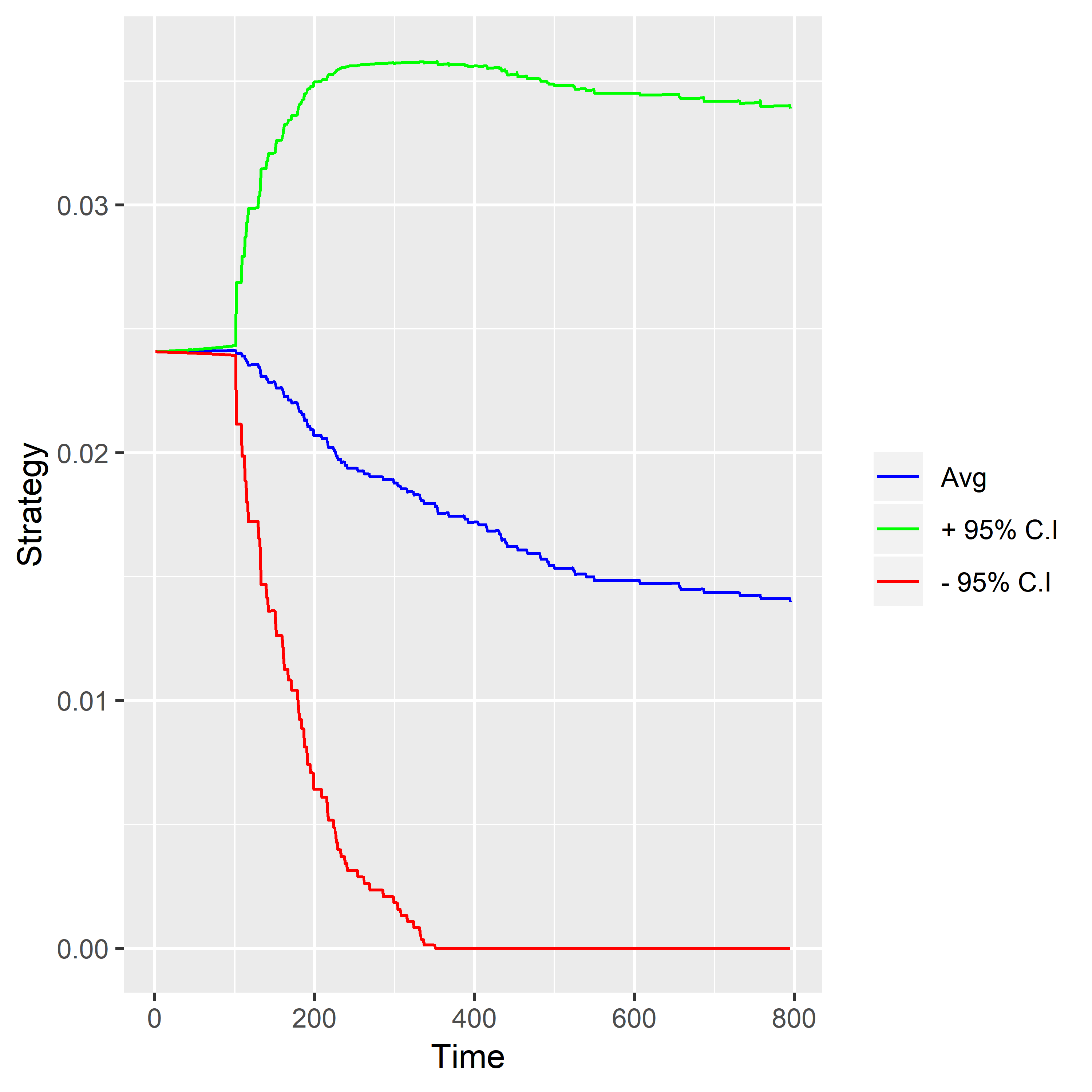}
\caption{$\beta = 0.0005$ }
  \label{fig:30}
\end{subfigure}
\begin{subfigure}{.6\textwidth}
  \includegraphics[width=0.95\textwidth]{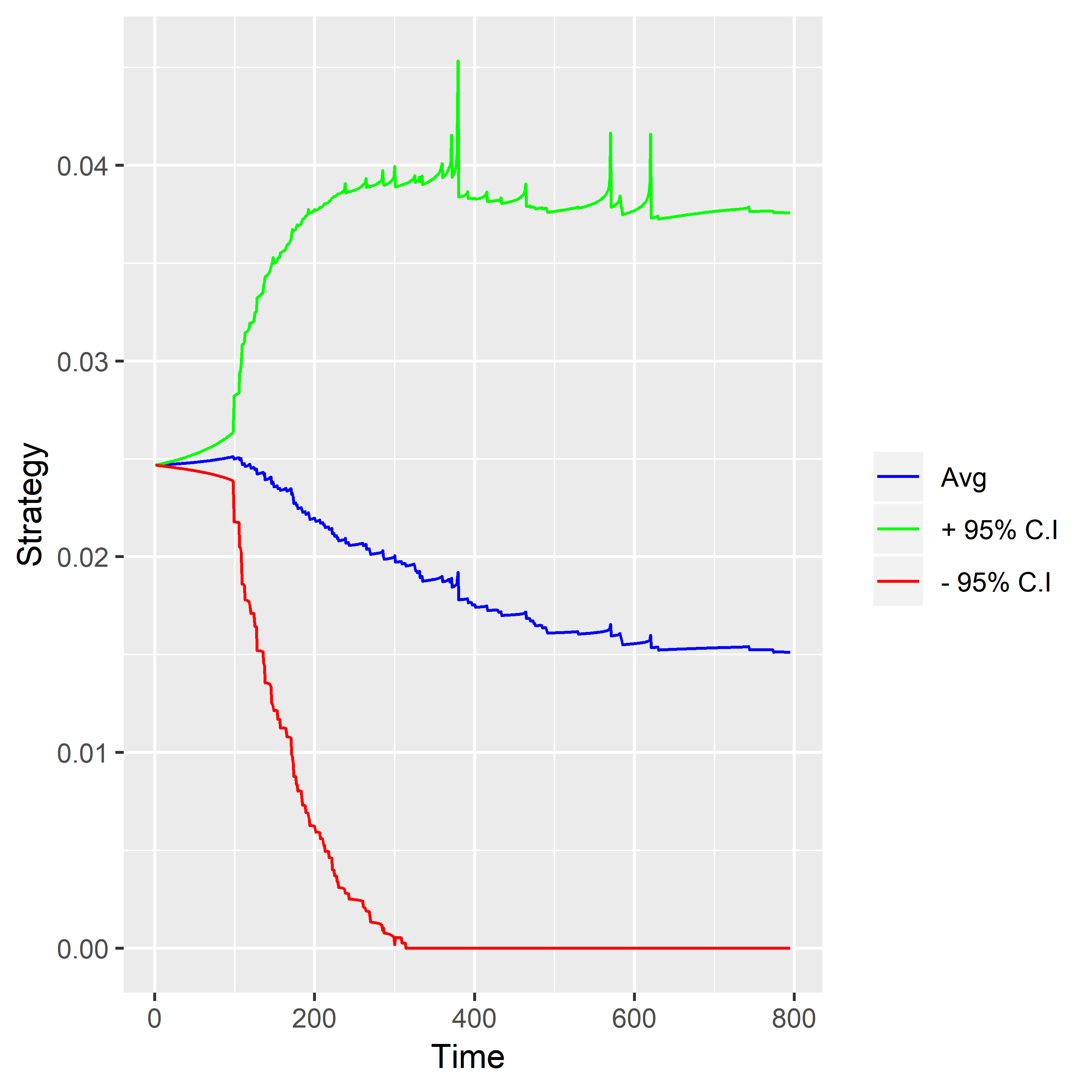}
   \caption{$\beta = 0.002$ }
  \label{fig:32}
\end{subfigure}
\caption{Average optimal strategy and its corresponding 95\% interval r = 0.00014, $\psi$ = 0.6, $\psi^{0}$ = 0.8, $\gamma$ = 0.3}
\label{fig:test}
\end{figure}

If we plot the wealth for all the three values of $\beta$ we can see that for all the cases the wealth accumulates in the same way, [refer figure (\ref{fig:21_3}, \ref{fig:31} and \ref{fig:33})]. So we can conclude that change in the discounting time period won't vary the level of wealth accumulation.
\begin{figure}[H]
\begin{subfigure}{.6\textwidth}
  \includegraphics[width=0.95\textwidth]{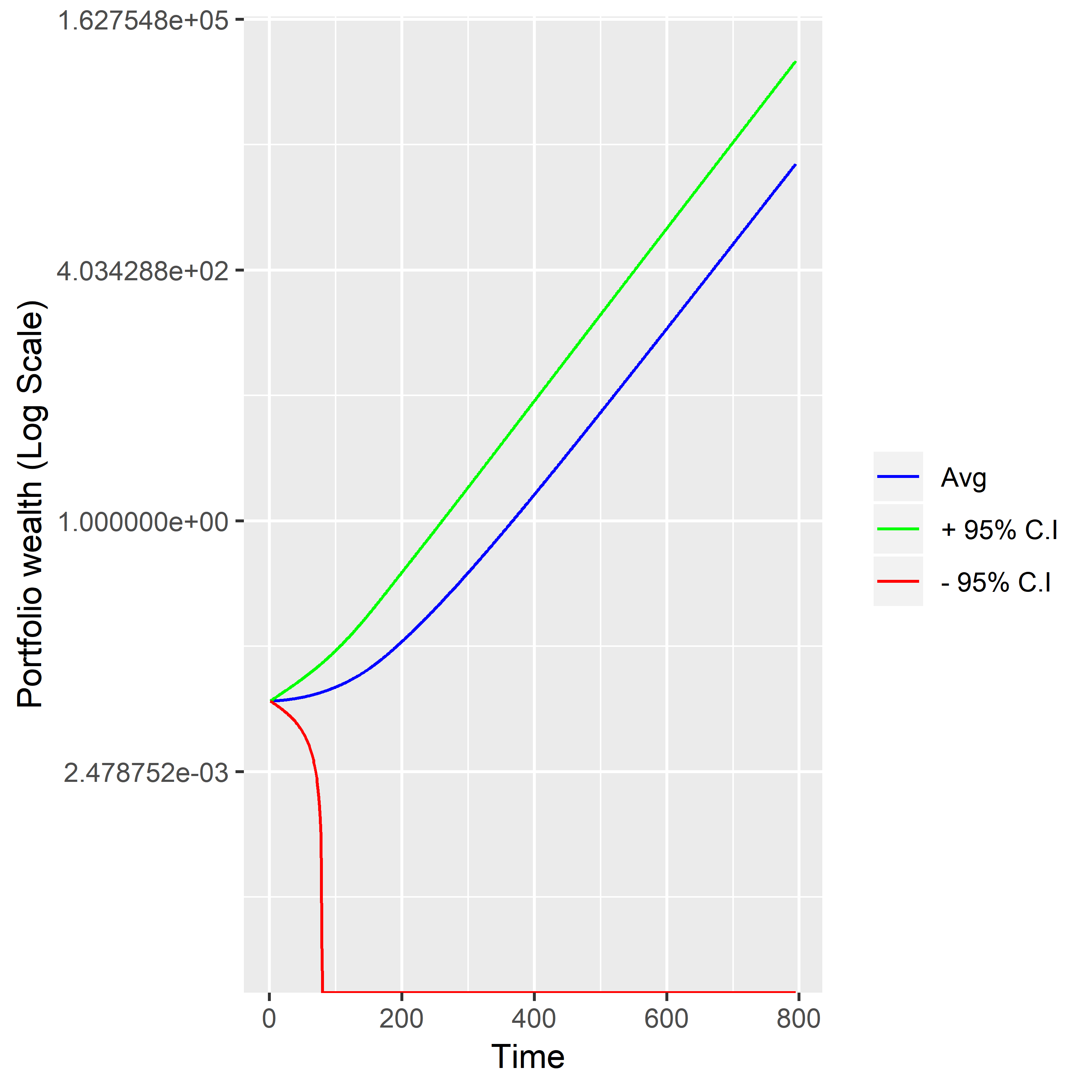}
   \caption{$\beta = 0.001$ }
  \label{fig:21_3}
\end{subfigure}
\begin{subfigure}{.6\textwidth}
  \includegraphics[width=0.95\textwidth]{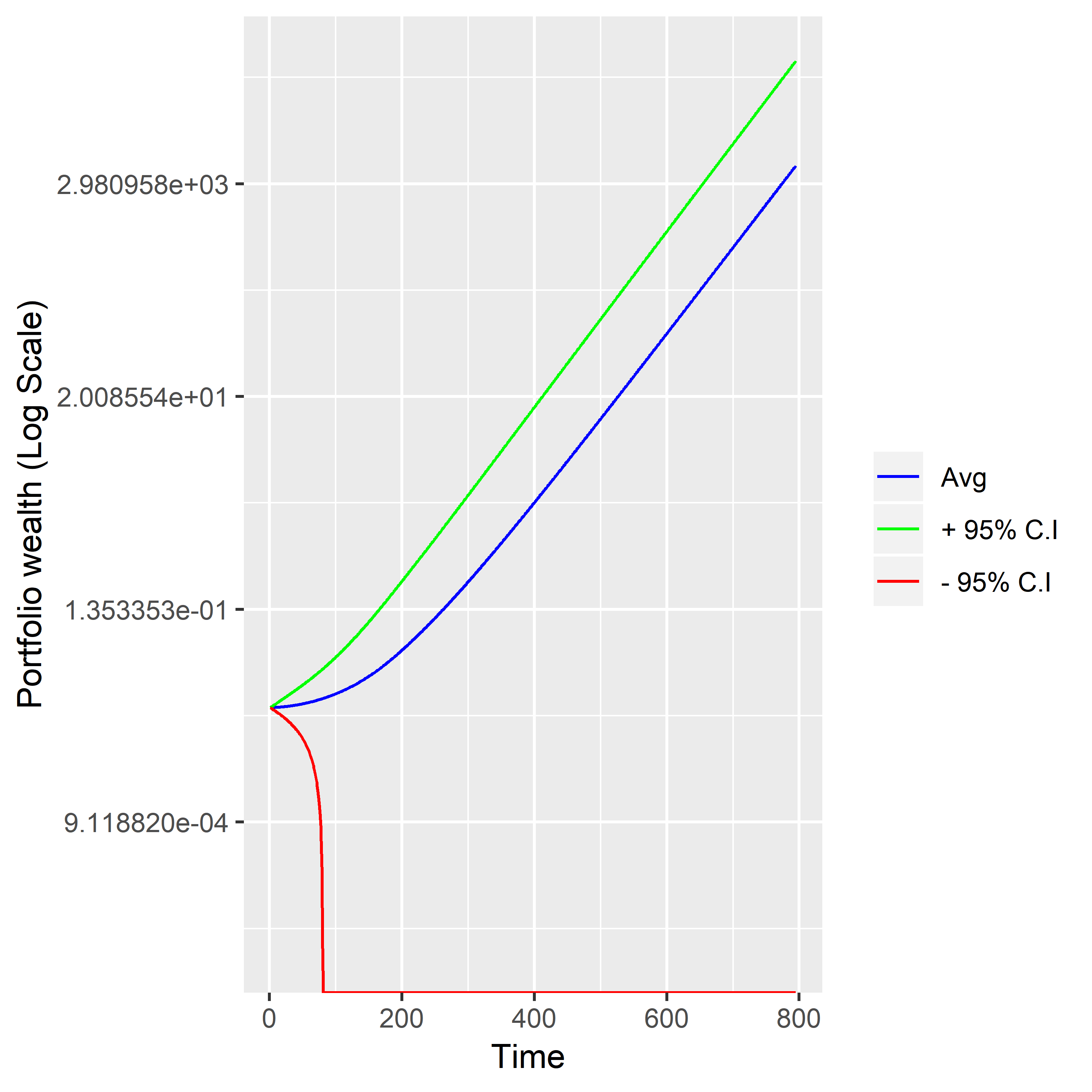}
\caption{$\beta = 0.0005$ }
  \label{fig:31}
\end{subfigure}
\begin{subfigure}{.6\textwidth}
  \includegraphics[width=0.95\textwidth]{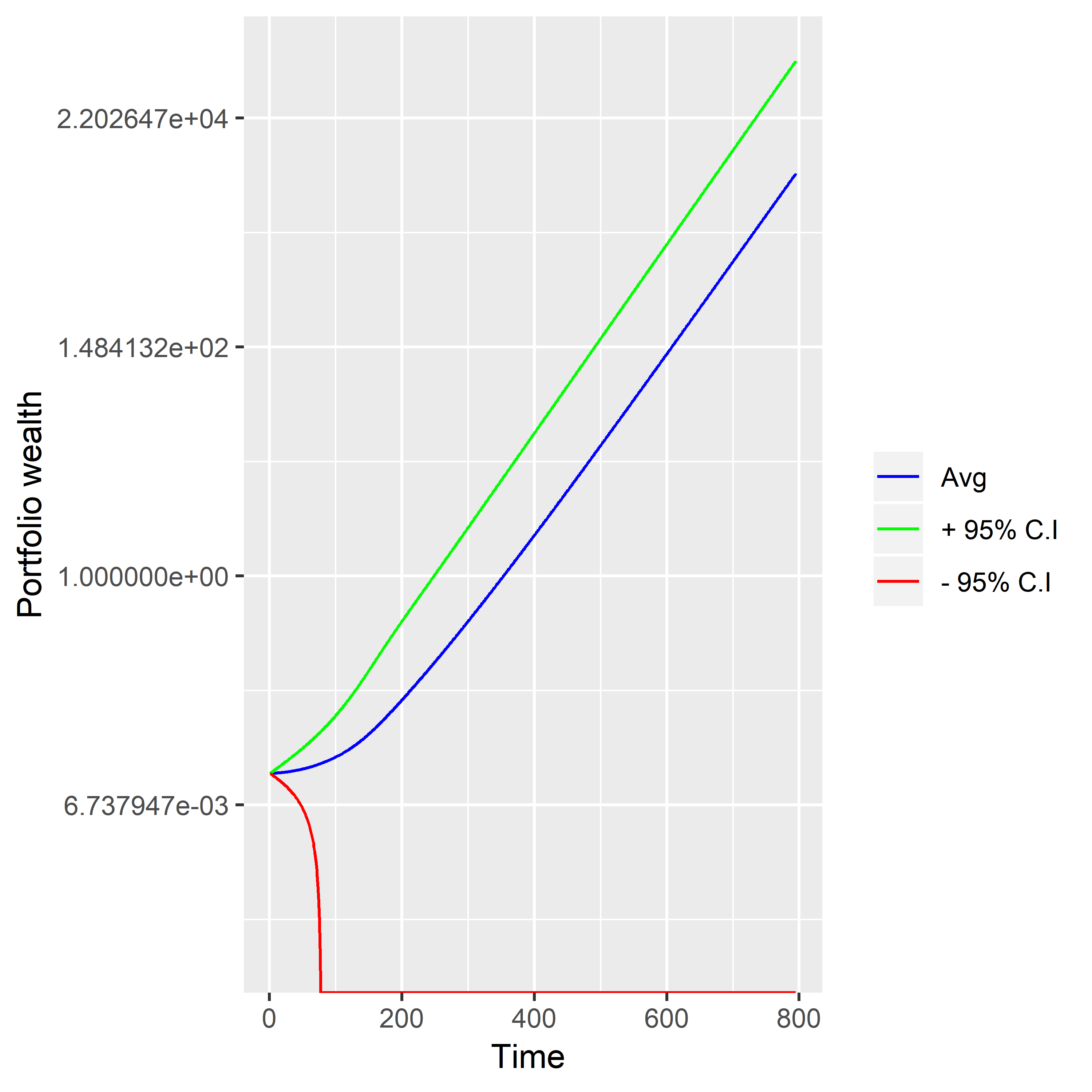}
   \caption{$\beta = 0.002$ }
  \label{fig:33}
\end{subfigure}
\caption{Average wealth and its corresponding 95\% interval r = 0.00014, $\psi$ = 0.6, $\psi^{0}$ = 0.8, $\gamma$ = 0.3}
\label{fig:test}
\end{figure}

\end{itemize}
\begin{center}
\textbf{Figure 7 and Figure 8 should be placed here}
\end{center}
\section{Concluding Remarks}

In this paper we have considered the optimisation problem for an investor whose portfolio consists of a single risky asset and a risk free asset. She wants to maximize her expected utility of the portfolio, in continuous time, subject to managing the Value at Risk (VaR) assuming a heavy tailed distribution of the stock price's return. We used the fact that the quantiles of the heavy tail distribution asymptotically follows normal distribution, allowing us to formulate the stochastic differential equation for the quantiles conveniently. The candidate utility function considered here is the power utility function with a constant relative risk aversion. But we discuss that similar results can be obtained, at the cost of some algebraic tedium, for a more general class of utility functions with constant risk aversion 
(e.g. logarithmic or exponential utility functions).

We use stochastic maximum principle to formulate the dynamic optimisation problem as in \citeauthor{Ref13}\cite{Ref13}. The equations which we obtain does not have any explicit analytical solution, so we look for accurate approximations to estimate the value function and optimal strategy. As our calibration strategy is non-parametric in nature, no prior knowledge on the form of the distribution function is needed.

We provided a detailed empirical illustration based on parameter values calibrated from a real life data and a range of choices for the subjective parameters. Our results show close concordance with financial intuition.
As this kind of a risk tolerance based portfolio optimization exercise has not been attempted in continuous time before, our results are expected to add to the arsenal of the portfolio managers who deals in high frequency trading.




\end{document}